\documentclass[final,5p,times]{elsarticle}
\usepackage{lineno,hyperref}
\modulolinenumbers[5]
\usepackage{amsmath,amssymb,amsfonts}
\usepackage{algorithmic}
\usepackage{graphicx}
\DeclareGraphicsExtensions{.pdf,.jpeg,.png}
\usepackage{subfigure}
\usepackage{epsfig}
\usepackage{booktabs} 
\usepackage{multirow}
\usepackage{array}
\usepackage{txfonts}
\usepackage{dirtree}
\usepackage{epstopdf}
\usepackage{caption,setspace}
\usepackage{wrapfig}
\usepackage{graphics}
\usepackage{diagbox}
\usepackage{float}
\usepackage{graphicx,color}
\graphicspath{{figure/}}
\usepackage{lipsum} 
\usepackage{float}
\usepackage{multirow}
\usepackage[table,xcdraw]{xcolor}
\usepackage{makecell,rotating,multirow,diagbox}
\usepackage{hyperref}
\usepackage{stfloats}
\usepackage{txfonts}
\journal{Journal of \LaTeX\ Templates}

\biboptions{sort&compress}
\begin{document}
\begin{frontmatter}

  \title{High-Capacity  Reversible Data Hiding in Encrypted 3D Mesh Models Based on Multi-MSB Prediction}
  
  \author[1]{Wan-Li Lyu}
  \author[1]{Lulu Cheng}
  \address[1]{Anhui Province Key Laboratory of Multimodal Cognitive Computation, School of Computer Science and Technology, Anhui University, 230601, P.R.China}
  \author[2]{Zhaoxia Yin\corref{mycorrespondingauthor}}
  \address[2]{School of Communication \& Electronic Engineering, East China Normal University,Shanghai 200241, China}
  \ead{Zhaoxia.edu@gmail.com}
  \cortext[mycorrespondingauthor]{Corresponding author}
  \begin{abstract}
    The interaction of massive data in the network world poses an enormous threat to data security. Reversible data hiding in encrypted domain (RDH-ED) technology ensures the security of data storage. Extensive works of RDH-ED take digital images as the cover mediums. With the advancement of multimedia technology, three-dimensional (3D) models are widely used on the network. In recent years, 3D models have also been used by researchers as cover mediums for RDH-ED technology. However, it is still challenging to improve the embedding rate under the premise of ensuring reversibility. This paper proposes an RDH-ED method to improve the embedding rate significantly, and the 3D mesh model can be restored after data extraction. Firstly, the vertices of the 3D mesh models are classified into ``embedded set'' and ``prediction set'' based on odd-even property of indices. Based on the high correlation of the two sets, multiple most significant bit (multi-MSB) prediction is adopted to calculate room redundancy. Arithmetic coding compresses auxiliary information generated during the reserved room phase to further free up space. In addition, the correlation between two sets is used to recover vertices that were modified due to data hiding. Experimental results show the proposed method greatly improves the embedding rate compared with state-of-the-art methods and guarantees model recovery with high quality.
  \end{abstract}
\end{frontmatter}


\section{Introduction}
\label{sec::introduction}
\par Data hiding~\cite{barton1997method} is a security technique that uses the information redundancy present in carrier data to carry additional data for privacy protection, secret communication, digital signature, copyright authentication, etc. In some special application scenarios, for example, in medical diagnostics and military applications, it is critical to restoring the marked media to the original carrier media after the additional data has been extracted~\cite{puech2008reversible,zhang2011reversible}. With the popularity of cloud computing, the data security problem in the big data environment is very serious. Outsourced storage in the cloud has become one of the solutions to solve the data security in the cloud environment. The outsourced storage can be managed by adding some additional data such as file types, annotation information, etc. to the digital media, but the cloud service provider has no right to cause permanent distortion to the digital media when embedding the additional data. Reversible data hiding (RDH) technology solves the problem of permanent distortion of digital media caused by data embedding. Typical RDH methods~\cite{ni2006reversible,jia2019reversible,gao2020reversible,tian2003reversible,fan2021reversible,celik2006lossless,zhang2013recursive} are mainly based on (1) histogram shifting~\cite{ni2006reversible,jia2019reversible,gao2020reversible}, (2) difference expansion~\cite{tian2003reversible,fan2021reversible}, and (3) lossless compression~\cite{celik2006lossless,zhang2013recursive}. RDH methods for digital images have been studied for many years, and theoretical optimal solutions have been achieved. However, for other digital carriers, such as audio, video, and 3D models, RDH method research is still in its infancy. 
\par 3D model is an emerging digital media after image, audio, and video, which is widely used in medical organ production, video game production, building structure display, and other fields. Research on RDH techniques for 3D models meets the urgent need for multimedia data security in complex and diverse social environments. The existing RDH methods for 3D models are divided into four categories: spatial domain, transformed domain, compressed domain, and encrypted domain. The first category method~\cite{wu2005reversible,wu2008reversible,zhang2019reversibility,jiang2018reversible} embeds data by modifying the vertex coordinates of the 3D models. In~\cite{wu2005reversible}, Wu \emph{et al.} proposed to embed the data into the 3D mesh model by modulating the distance from the mesh surface to the center of mass of the mesh. In ~\cite{wu2008reversible,zhang2019reversibility}, the classical image prediction error expansion method is extended to the 3D model, and the difference between the predicted vertex coordinates and the original vertex coordinates are used to embed secret data. In~\cite{jiang2018reversible}, Jiang \emph{et al.} proposed an RDH method for 3D mesh models based on the optimal three-dimensional prediction-error histogram modification with recursive construction coding.
The second category method embeds secret data in the frequency domain space of the 3D models. In literature~\cite{luo2006reversible}, the eight vertices are first divided into clusters, and the seed vertices are randomly selected. Then the watermarked bits are modulated in the integer DCT domain using an efficient highest frequency coefficient correction technique. The third category method~\cite{sun2006reversible,lu2007high,bhardwaj2022efficient} uses compression algorithms to compress and encode 3D model and then embed the secret data.~\cite{sun2006reversible} and~\cite{lu2007high} compressed the vertex data of a mesh with vector quantization techniques and embedded secret data by modifying the prediction rules during the vector quantization. In~\cite{bhardwaj2022efficient}, hierarchical absolute moment block truncation coding is used to compress the mesh blocks and then embed secret data in the compressed domain. The fourth category is 3D models RDH in encryption domain (RDH-ED). RDH technology helps cloud servers to solve the security problem of stored data. But, in outsourced storage applications, the content of digital multimedia can be easily copied and modified, resulting in user privacy leakage. Encryption is necessary to protect content privacy. RDH-ED technology was first proposed for digital images~\cite{zhang2011separable,wu2014high,hong2012improved,qin2018separable,ma2013reversible,zhang2014reversibility,yin2021reversible} and was classified into two categories according to the different encryption sequence: vacating room after encryption (VRAE)~\cite{zhang2011separable,wu2014high,hong2012improved,qin2018separable} and reserving room before encryption (RRBE)~\cite{ma2013reversible,zhang2014reversibility,yin2021reversible}. VRAE-based framework mainly utilizes part of spatial correlation of encrypted image pixels to embed additional data. However, the VRAE-based framework has its limitations; that is, encryption operation destroys the correlation pixels, which may lead to bit error rate in data extraction or image recovery. Ma \emph{et al.}~\cite{ma2013reversible} first proposed an RDH method for RRBE. The content-owner uses the pixel correlation of the carrier image to reserve embeddable room and encrypts the image. This method makes data extraction and image restoration without any error. Considering 3D models as digital media, the researchers proposed to encrypt vertex coordinates of 3D model first and then embed the secret data.
 \begin{figure*}[bp]
\centering
  \includegraphics[width=17cm,height=5.6cm]{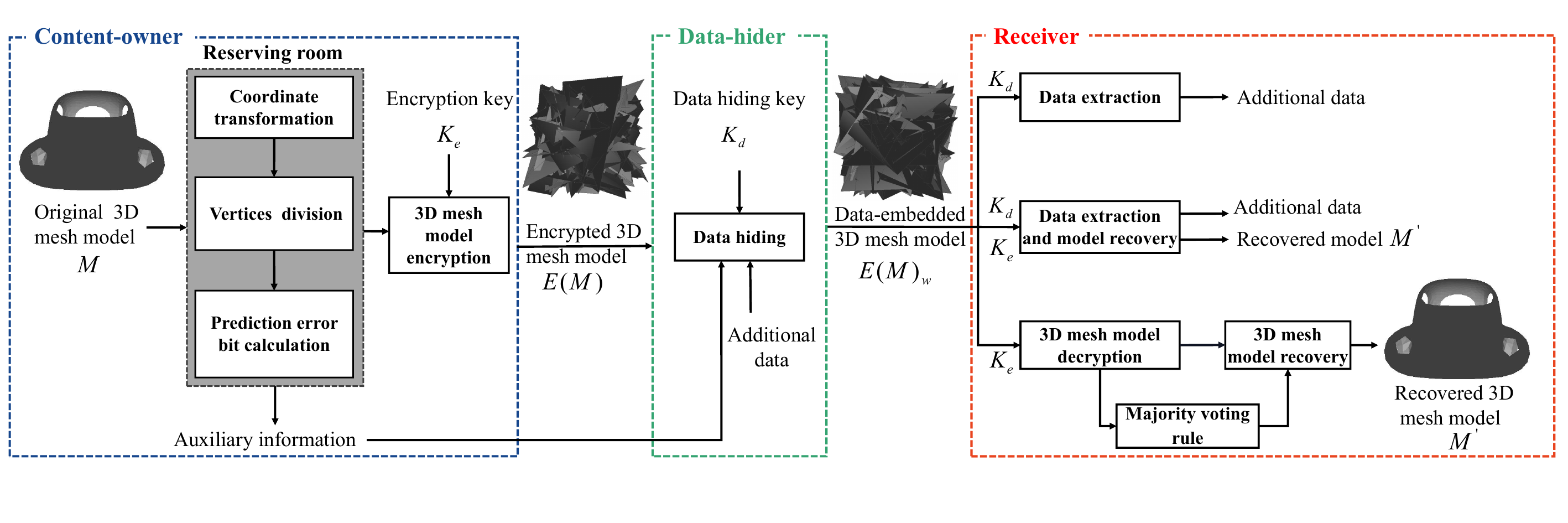}
\begin{center}
  \caption{The framework of the proposed method.}
  \label{figure_1}
\end{center}
\end{figure*}
In 2017, Jiang \emph{et al.}~\cite{jiang2017reversible} proposed a method of VRAE for 3D models and extended RDH in encrypted images to encrypted 3D models for the first time. This method inverts the least significant bits (LSBs) of vertex coordinate to embed 1-bit data, then define a smoothing function to measure fluctuations to evaluate the smoothness of local areas to locate and extract data. In~\cite{shah2018homomorphic}, Shah \emph{et al.} implemented a two-tier embedding strategy using the self-blinding property of the Paillier cryptosystem and histogram translation. Compared with~\cite{jiang2017reversible}, the embedding rate is greatly improved by double-layer embedding. Applying Paillier cryptosystem enhances the security of model content and data, but neglects that can lead to vertex coordinate value expansion and low computational efficiency, which limits the application of this method in real-world scenarios. In~\cite{van2021homomorphic}, a new two-layer RDH method for 3D models based on Paillier encryption is proposed. The method first groups vertices into blocks and processes vertex coordinates through format conversion, then sets the specified bits in the block to solve the problem of vertex coordinate expansion. However, this method can not restore the original model losslessly, and the encryption method also causes low computational efficiency. In~\cite{tsai2020separable}, an efficient RDH method was designed by using space coding and spatial subdivision techniques. The embedding rate is higher than in the previous works, but if the threshold value is improperly selected, there will be errors in data extraction. The method of RRBE can further utilize the spatial correlation of the original model. Based on ~\cite{jiang2017reversible}, Xu \emph{et al.}~\cite{xu2021separable} believed that the most significant bit (MSB) has more redundant room and proposed an RRBE-based RDH method in 3D mesh models based on MSB prediction and integer-mapping. Their method explores the high correlation of MSB and obtained better performance than~\cite{jiang2017reversible}. To further utilize the redundant room of vertices, 
Yin \emph{et al.}~\cite{yin2021separable} extended MSB to mutil-MSB to obtain more embedding room. The data-hider first sets the length of data embedding. When the embeddable length of a vertex is less than the set value, the vertex is classified as the vertex with wrong prediction and it cannot be used for embedding additional data, so the redundant room of the model cannot be fully utilized. We propose a method to exploit the redundancy of vertices further. The main contributions of the proposed method are as follows:
\par 1) In this paper, considering the different embeddable length of each vertex, a label list is used to identify the embeddable room of each vertex and additional data is embedded in each vertex adaptively according to the embeddable capacity of each vertex, which further exploits the redundancy of vertices and greatly improves the embedding rate. In addition, the proposed method can restore the carrier model losslessly by selecting an appropriate truncation coefficient $u$.
\par 2) In~\cite{yin2021separable}, all vertex coordinates are traversed in ascending order according to the topological structure of the triangular face. This method yields fewer embeddable vertices than non-embeddable vertices. Different from the way of dividing sets in literature~\cite{yin2021separable}, this paper divides ``embedded set'' and ``prediction set'' according to the odd-even property of indices, which not only guarantees that half of vertex coordinates can be used to free up room, but also ensures the independence of vertices in two sets so that the modified vertices can be recovered during the model recovery phase.
\par 3) Arithmetic coding is used to compress auxiliary information with sparse features, which frees up more room for additional data embedding. 
\par The rest of this paper is organized as follows: Section~\ref{sec::Proposed method} introduces the detailed process of our method. In Section~\ref{sec::Experimental}, experimental results and analysis are presented. This paper presents conclusion of the proposed method and discussion of future work in Section~\ref{sec::Conclusion}.
\section{Proposed method}
\label{sec::Proposed method}
\par The framework contains three parts: (1) The content-owner reserves room for data hiding and encrypts original 3D mesh model; (2) The data-hider embeds additional data in the redundant room of vertex coordinates; (3) The receiver extracts the additional data and restores the original contents of the 3D mesh model. The framework to describe the operation process of our method is shown in Fig.\ref{figure_1}. Section~\ref{subsec::vacate room} describes the process of generating room before encryption. 3D mesh model encryption process appears in Section~\ref{subsec::encryption}. In Section~\ref{subsec::data embedding}, procedures of data embedding are given. Finally, the process for data extraction and model recovery by receivers with different keys are described in Section\ref{subsec::data extraction}.
\begin{table*}[bp]
\setlength{\belowcaptionskip}{0.2cm}
\begin{center}
\renewcommand{\arraystretch}{1.2}
\caption{Data structure of triangular index\cite{jiang2017reversible}}
\label{tab_1}
\centering
\begin{tabular}{cccccc}
\toprule[1pt]
\multicolumn{4}{c}{Vertex coordinate list}&\multicolumn{2}{c}{Triangular face list}\\
\cmidrule(r){1-4}  \cmidrule(r){5-6}
Index\ of\ vertex&$x$-axis coordinate&$y$-axis coordinate&$z$-axis coordinate&Index\ of\ face&Elements\ of\ face \\
\midrule
1&$v_{1, x}$&$v_{1, y}$&$v_{1, z}$&1&(16,0,1)\\
2&$v_{2, x}$&$v_{2, y}$&$v_{2, z}$&2&(2,1,16)\\
3&$v_{3, x}$&$v_{3, y}$&$v_{3, z}$&3&(3,2,17)\\
4&$v_{4, x}$&$v_{4, y}$&$v_{4, z}$&4&(4,3,18)\\
5&$v_{5, x}$&$v_{5, y}$&$v_{5, z}$&5&(5,4,19)\\
6&$v_{6, x}$&$v_{6, y}$&$v_{6, z}$&...&...\\
...&...&...&...&16&(6,5,20)\\
17&$v_{17, x}$&$v_{17, y}$&$v_{17, z}$&17&(7,6,21)\\
18&$v_{18, x}$&$v_{18, y}$&$v_{18, z}$&...&...\\
...&...&...&...&...&...\\
31&$v_{31, x}$&$v_{31, y}$&$v_{31, z}$&...&...\\
...&...&...&...&...&...\\
\bottomrule[1pt]
\end{tabular}
\end{center}
\end{table*}
\subsection{Vacating room before encryption}
\label{subsec::vacate room}
\subsubsection{Vertex coordinate transformation}

\par In this paper, the decimal floating-point vertex coordinates of original 3D mesh model \emph{M} are converted to binary form according to the corresponding rule, the detailed operation process is explained as follows: 3D mesh model studied in this paper is a proper approximation of a real three-dimensional object, which is made up of interconnected triangles. 3D mesh model is represented by two types of information: geometric and topological. Geometric information represents vertices position and size of mesh in space, and topological information stand for the connectivity between the various parts of the mesh. 3D data is stored in various file formats (OFF, OBJ, MD2, etc.) in different application scenarios. This paper discusses the 3D mesh models in the OFF format. A polygonal mesh model consists of two important sets of parameters: the set of vertices $V=\left\{v_{1}, v_{2}, \ldots, v_{p}\right\}$, $p$ is the number of vertices, and the set of surfaces $F=\left\{f_{1}, f_{2}, \ldots, f_{q}\right\}$, $q$ is the number of face. A triangular face consists of three vertices. Table~\ref{tab_1} shows the format of the corresponding file. The vertex coordinates of the uncompressed 3D mesh model are represented as 32-bit floating-point numbers, but the valid bits are 6 bits. For example, $v=\{0.295122,0.338337,0.103678\}$. In~\cite{deering1995geometry}, Deering suggested that most applications of 3D models should be performed at an acceptable level of accuracy, without the need to operate on every bit of floating point coordinates, and gives a method to compress the vertex data. In this paper, the truncation function is used to normalize the position of all vertices with the boundary box based on coordinate axes for parts below the accuracy level, and then the position coordinates are uniformly quantified to the bit level, so vertex coordinate of model can be empirically expressed as integers between 0 and $2^{u-1}$, where $u\in[1,33]$, and $u$ is the truncation coefficient; vertex coordinate $v_{i,j}$, where $i\in[1,N]$, $N$ is the number of vertices, and $v_{i,j}$ is normalized to 
\begin{eqnarray}
v_{i, j}^{\prime}=\left\lfloor v_{i, j} \times 10^{u}\right\rfloor, \quad j=x, y, z.
\end{eqnarray} 
\begin{figure}[!t]
\begin{center}
\vspace{-0.6cm}
  \includegraphics[width=8cm,height=4cm]{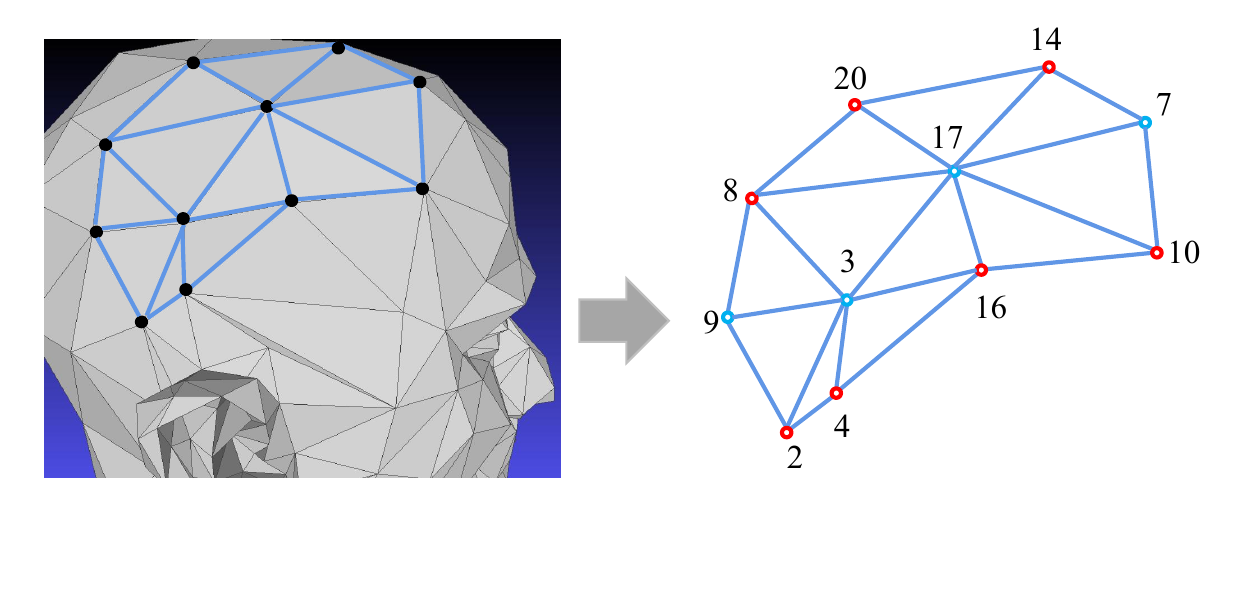}
  \caption{A 3D mesh model (Mannequin) topology visualization. }
  \label{fig_2}
\end{center}
\end{figure}
\begin{figure*}[!t]
\centering
  \includegraphics[width=17.8cm,height=7.5cm]{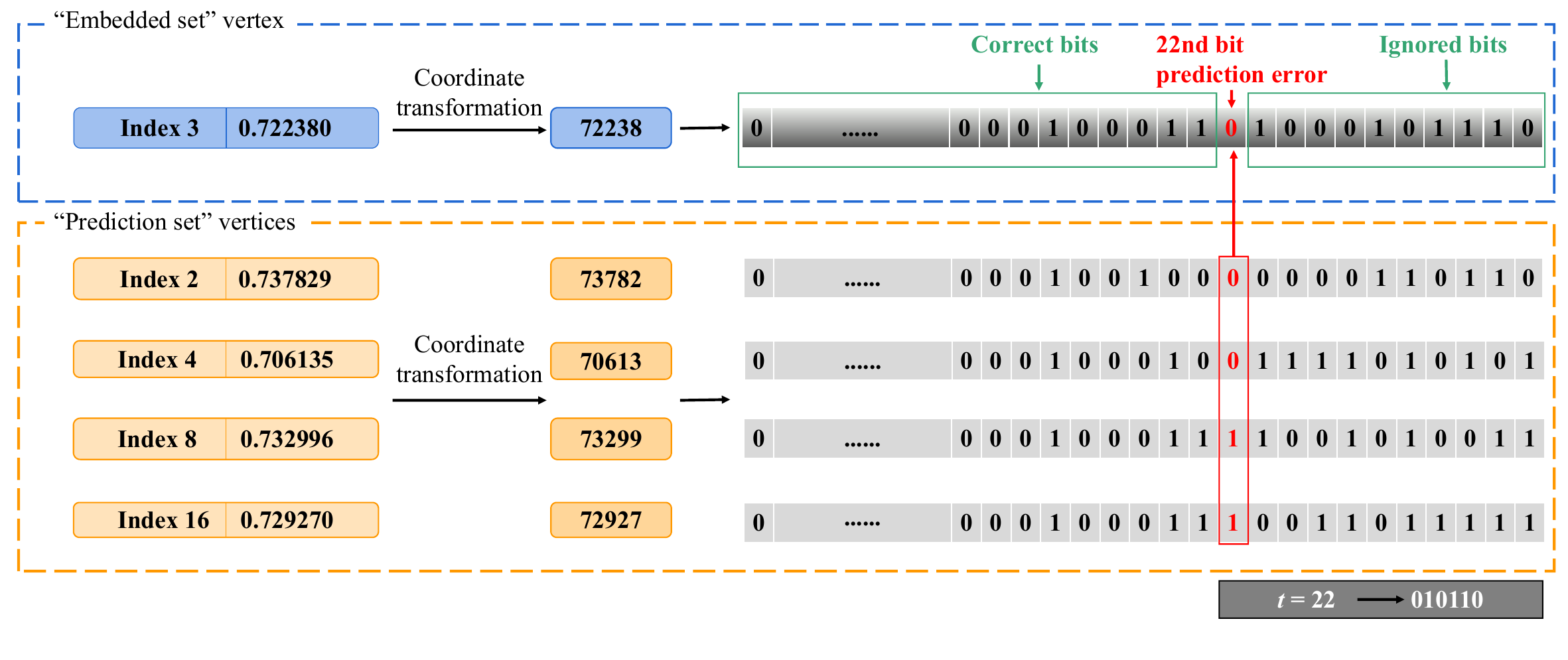}
\begin{center}
  \caption{ Prediction error bit detection.}
  \label{fig_3}
\end{center}
\end{figure*}
In the model recovery phase, integral coordinates $v_{i, j}^{\prime}$ need to be converted to original floating-point form using
\begin{eqnarray}
\widetilde{v}_{i, j}=v_{i, j}^{\prime} / 10^{u}, \quad j=x, y, z.
\end{eqnarray}
According to the above analysis, truncation coefficient $u$ affects the precision of coordinates, and then affects the embedding rate and model recovery quality in the process of coordinate transformation. $u$ determines the bit length $l$ of the vertex coordinates, the specific rules are summarized as
\begin{eqnarray}
\label{test_1}
l= \begin{cases}8, & 1 \leq u \leq 2 \\ 16, & 3 \leq u \leq 4 \\ 32, & 5 \leq u \leq 9 \\ 64, & 10 \leq u \leq 33 .\end{cases}
\end{eqnarray}
\subsubsection{Vertices division}
Content-owner reserves redundant room for data embedding by exploring correlations between vertices. To ensure original 3D mesh models can be restored perfectly after data extraction, vertices are classified as ``embedded set'' $\mathcal{S}_{e}$ and ``prediction set'' $\mathcal{S}_{p}$ according to the odd-even property. That is, add the vertices with odd index to one set, and the vertices with even index to another set. To better explain our method, the local area of ``Mannequin'' model is shown in Fig.~\ref{fig_2}. Blue dots with odd index represent vertices in $\mathcal{S}_{e}$ and red dots with odd index represent vertices in $\mathcal{S}_{p}$. $\mathcal{S}_{e}$ is the set used to embed the additional data, and $\mathcal{S}_{p}$ is used to recover the vertices in $\mathcal{S}_{e}$. Note that the coordinates of the vertices in the $\mathcal{S}_{e}$ will be offset due to data embedding. However, the state of vertices in $\mathcal{S}_{p}$ remains unchanged during the data embedding; otherwise, the original model cannot be restored. In Fig.~\ref{fig_2}, for example, $(3,8,9)$ is the first triangular face to be traversed, so add vertex 3 with odd-index to $\mathcal{S}_{e}$, and add the even-index vertices adjacent to vertex 3 to $\mathcal{S}_{p}$, namely $\mathcal{S}_{e}=\{3\}$, $\mathcal{S}_{p}=\{8\}$, and its 1-ring neighborhood contains vertices 2, 4, 8, 9, 16, and 17. The vertices with even indexes adjacent to vertex 3 are added to $\mathcal{S}_{p}$, that is $\mathcal{S}_{p}=\{2,4,8,16\}$.  Similarly, vertex $8$, $10$, $14$, $16$, and $20$ are used to recover vertex 17, that is $\mathcal{S}_{e}=\{3 \cup 17\}$, $\mathcal{S}_{p}=\{2,4,8,10,14,16,20\}$, one by one until all the vertices are added sets. Therefore, the change of the vertex coordinates in ``embedded set'' does not affect the vertex position in ``prediction set''. This ensures that the prediction result of the ``prediction set'' about the vertex coordinates of the ``embedded set'' is consistent in the process of data extraction and data hiding. Therefore, the correlation between the two sets can be used to recover the vertices in ``embedded set''. The following section describes the detailed operation of reserving room.
\subsubsection{Prediction error bit detection scheme}
\label{subsub::prediction error}
\par Depending on the relative proximity of adjacent vertices, MSBs of vertex coordinate have a higher correlation than LSBs. So we explore the correlation of MSBs of 1-ring neighborhood vertices to embed data. Fig.~\ref{fig_3} shows an example of the process of prediction error bit detection, The coordinates of vertex $3$ and its neighborhood vertices with even-index are expressed in binary form by vertex coordinate transformation, $u$ is equal to 5 in this process. In a local area, the central vertex and its adjacent vertex are closer in space, and the coordinate correlation utilization rate is higher. Content-owner counts the number of `0' bit and `1' bit number of the multi-MSB of vertex 2, 4, 8, 16. The predicted value can be obtained based on the majority voting rule in the voting mechanism. The predicted value is calculated as
\begin{eqnarray}
p_{i, j}^{k}=\left\{\begin{array}{ll}
1, & N_{1, k} \geq N_{0, k} \\
0, & N_{0, k}>N_{1, k}
\end{array} \quad k=0,1, \ldots, l-1\right.
\end{eqnarray}
where $p_{i,j}^{k}$ is the $k$-th predicted value of vertex coordinate. $N_{1, k}$ is the number of `1'  in the $k$-MSB of vertices in $\mathcal{S}_{p}$. $N_{0, k}$ is the number of `0' in the $k$-MSB of vertices in $\mathcal{S}_{p}$. If the number of `1' bit predicted is greater than or equal to the number of `0' bit, then the predicted value the MSB of the current vertex coordinate in $\mathcal{S}_{e}$ is 1; otherwise, predicted value is 0. Assuming that label $t$ records the bit that predicts error. Compare each bit of $v_{i, j}^{t-M S B}$ and $p_{i, j}^{t-M S B}$ sequentially from MSB to LSB until a certain bit is different,
\begin{eqnarray}
\begin{aligned}
&\underset{t}{\arg \max }\ v_{i, j}^{t-M S B}=p_{i, j}^{t-M S B},\quad t=1,2, \ldots, l\\
&\text { subject to } v_{i, j}^{t-M S B} =\left\lfloor v_{i, j} / 2^{l-t}\right\rfloor \bmod 2,
\end{aligned}
\end{eqnarray}
where $v_{i,j}$ is integer coordinate of vertex, $v_{i, j}^{t-M S B}$ and $p_{i, j}^{t-M S B}$ are the $t$-MSB values of $v_{i,j}$ and $p_{i,j}$. 
As shown in Fig.~\ref{fig_3}, the 22-MSB of the binary coordinate of vertex 3 is inconsistent with the predicted value, that is, $t$ = 22. In our method, the $t$ value is first recorded with a fixed-length binary coding and represents the label as ``010110''. The label values of all vertices form a label list. The label list and its length are sent to the recipient as auxiliary information along with additional data for data extraction and model recovery. We use arithmetic coding to compress auxiliary information with sparse characteristics to reserve more bits for embedding additional data. 3D mesh models are carriers distributed in space. In a three-dimensional Cartesian coordinate system, vertices are represented by three-dimensional coordinates. Assuming that $r_1$, $r_2$, and $r_3$ represent the embeddable lengths of the $x$-axis, $y$-axis, and $z$-axis coordinate, respectively, the embeddable length of the vertex is $r=\min \left\{r_{1}, r_{2}, r_{3}\right\}$. 
\subsection{3D mesh model encryption}
\label{subsec::encryption}
\par The vertex coordinate transformation operation above convert all coordinates of vertices to binary bitstream. In our method, denote the bits of a component of binary form of a vertex coordinate as $b_{i, j, 0}$, $b_{i, j, 1}$, ..., $b_{i, j, k}$, where $1 \leq i \leq N$, $N$ is the number of vertices and $j \in\{x, y, z\}$. This means that
\begin{eqnarray}
b_{i, j, k}=\left\lfloor v_{i, j}^{\prime} / 2^{k}\right\rfloor \bmod 2, \quad k=0,1, \ldots, l-1 ,
\end{eqnarray}
$v_{i, j}^{\prime}$ is normalized vertex coordinate; $\lfloor * \rfloor$ is the lower bound function, $l$ is the binary bit length of vertex coordinate. The content-owner choose an encryption key $K_e$ to generate a random binary pseudo-random sequence $s_{i, j, k}$ to encrypt original model by
\begin{eqnarray}
e_{i, j, k}=b_{i, j, k} \oplus s_{i, j, k}, \quad k=0,1, \ldots, l-1,
\end{eqnarray}
where $e_{i, j, k}$ is a generated ciphertext, and $\oplus$ is XOR operation. The integral coordinates of the encrypted 3D mesh model $E(M)$ are computed as  
\begin{eqnarray}
v_{i, j}^{\prime \prime}=\sum_{k=1}^{l-1} e_{i, j, k} \cdot 2^{k},
\end{eqnarray}
encryption only changes the value of vertex coordinates, not the topology between vertices.
\subsection{Data hiding by mutil-MSB substitution}
\label{subsec::data embedding}
\par
\begin{figure*}[!t]
  \centering
 \includegraphics[width=14cm,height=10.5cm]{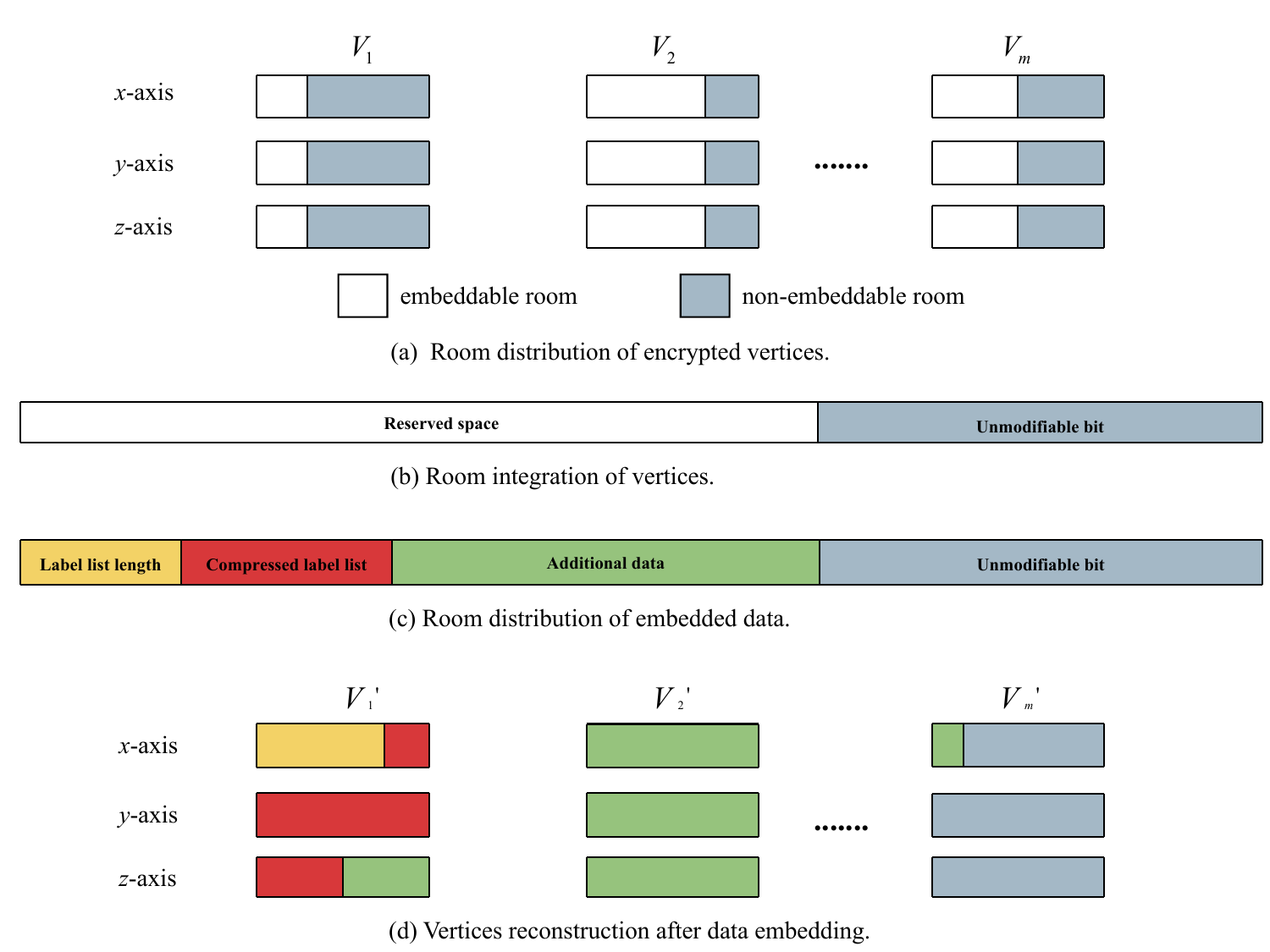}
\begin{center}
  \caption{The room distribution of data embedding process.}
  \label{fig_4}
\end{center}
\end{figure*}

 Through the above prediction error analysis, the ($t-1$) MSBs are embeddable bits that can be used to embed the additional data. The additional data is embedded according to the order of the vertices in ``embedded set'', generating data-embedded 3D mesh model $E(M)_{w}$. Convert the coordinates of encrypted vertices to binary form before embedding the data by 
\begin{eqnarray}
b_{i, j, k}^{\prime}=\left\lfloor v_{i, j}^{\prime\prime} / 2^{k}\right\rfloor \bmod 2, \quad k=0,1, \ldots, l-1,
\end{eqnarray}
$v_{i, j}^{\prime\prime}$ is the integral coordinates of the encrypted vertices. The room distribution of data embedding process is shown in Fig.\ref{fig_4}. In Section~\ref{subsub::prediction error}, we can obtain the label list used to record the embeddable length of the  original vertices in ``embedded set''. In Fig.~\ref{fig_4}(a), reserved room and non-embeddable room of $m$ encrypted vertices are shown in white and gray respectively. Integrate the room reserved for all ``embedded set'' vertices with the non-embeddable room is shown in Fig.~\ref{fig_4}(b). The label list recovers the vertices changed due to data embedding in the model recovery phase, it is necessary to embed the label list as auxiliary information along with the secret data into the vertices. The compressed label length and the compressed label list are embedded at the beginning of the reserved room to extract the label list accurately, where the compressed label list length is recorded with $\left\lceil\log _{2}\left(\left\lceil\log _{2} l\right\rceil\ * 3 * m\right)\right\rceil$ bit, $m$ is the number of vertices in ``embedded set'' and $l$ denotes the binary length of vertex coordinates. The remaining room for embedding additional data by mutil-MSB substitution, Fig.~\ref{fig_4}(c) shows the location of the auxiliary information and additional data storage. In the last step, after all the data is embedded, the binary bitstream needs to be converted into the floating-point form of the vertex coordinates, such as $V_{1}^{\prime}, V_{2}^{\prime}, \ldots \ldots, V_{m}^{\prime}$ as shown in Fig.~\ref{fig_4}(d).

\subsection{Data extraction and model recovery}
\label{subsec::data extraction}
\par Content-owner divides all vertices into ``embedded set'' and ``prediction set''. Similarly, the legal recipient first divides vertices of data-embedded 3D mesh model into the ``embedded set'' and ``prediction set''. The recipient can extract the length of the label list and the compressed label list from the data-embedded 3D mesh model, then locate and extract additional data according to label list. Whether the recipient can extract the data and restore the model depends on the different keys they own. Note that the length of the compressed label list is recorded in fixed-length bits and distributed to legitimate recipients.
\par 1) If the recipient has only the data hiding key $K_d$, the encrypted secret data can be extracted. The recipient uses the data hiding key $K_d$ to decrypt the secret data and then gets the plaintext data according to the auxiliary information. However, since there is no encryption key $K_e$, the original 3D mesh model cannot be reconstructed.
\par 2) If the recipient has only encryption key $K_e$, the model can be reconstructed. Encrypted models can be decrypted by using $K_e$. The coordinates of the vertices in ``embedded set'' are changed because of the additional data embedded. The vertices of the ``embedded set'' and ``prediction set'' are independent, and the states of the vertices in ``prediction set'' remain unchanged in the process of data embedding. So the recipient can exploit the correlation between vertices in 1-ring neighborhood to recover vertex coordinates in ``embedded set'' . As shown in Fig.~\ref{fig_3}, the ($t$-1) MSBs of vertex in ``embedded set'' are substituted with auxiliary information and additional data. ($t$-1) MSBs of vertex 3 can be accurately restored by performing majority vote operation on the vertices of 1-ring neighborhood. 
\par 3) If the recipient has both the data hiding key $K_d$ and the encryption key $K_e$, the recipient can not only extract additional data but also reconstruct the original 3D mesh model. The detailed steps are the same as above. To extract additional data accurately, the data must be extracted before model decryption.
\begin{figure*}[!t]
\centering
\subfigure[]{
\includegraphics[width=4cm]{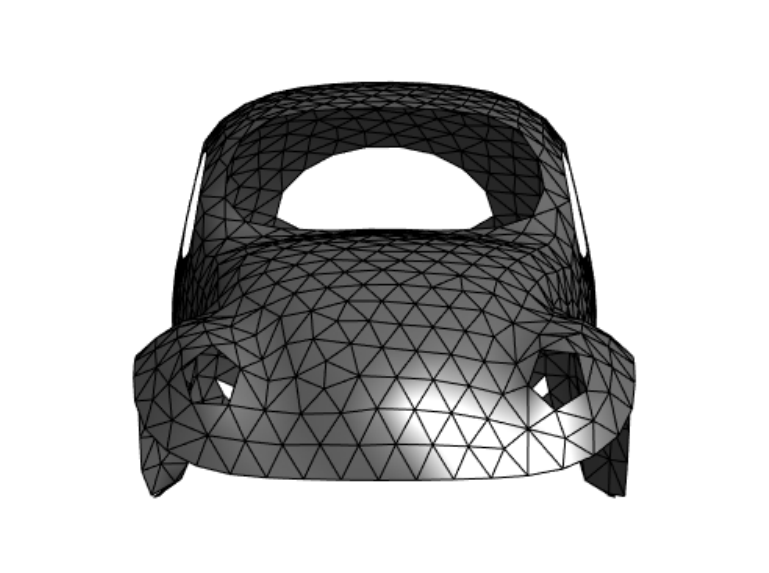}
}
\subfigure[]{
\includegraphics[width=4cm]{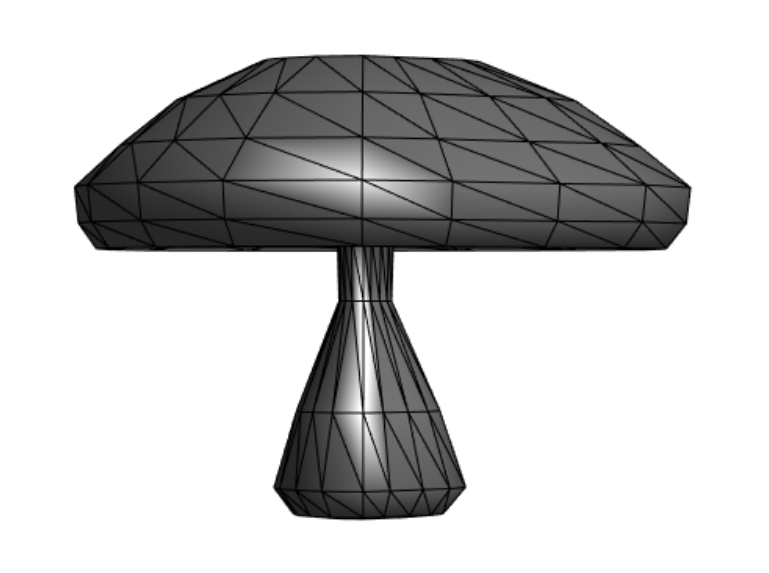}
}
\subfigure[]{
\includegraphics[width=4cm]{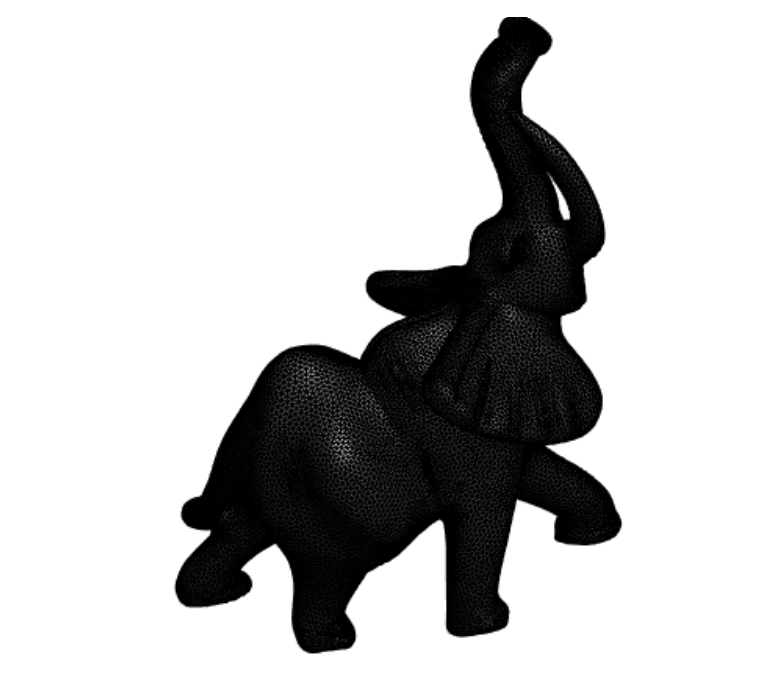}
}
\subfigure[]{
\includegraphics[width=4cm]{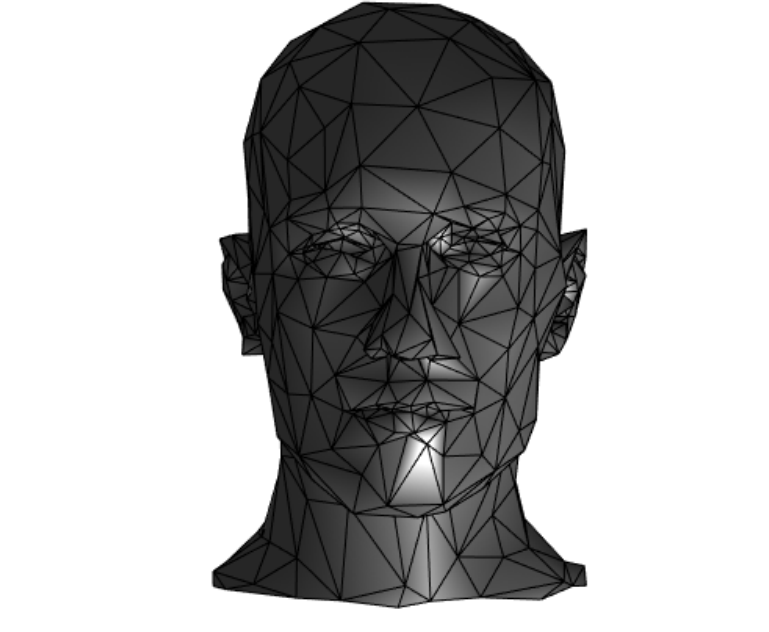}
}
\quad
\caption{\centering Test 3D mesh models: (a) Beetle;  (b) Mushroom;  (c) Elephant; (d) Mannequin.}
\label{fig_5}
\end{figure*}
\section{Experimental results}
\label{sec::Experimental}
\par To demonstrate the performance of our method, the corresponding analysis, and discussions are given in this section in terms of analysis of embedding rate (ER), visual quality of restored 3D mesh model and performance comparison with state-of-the-art methods. Finally, we also test and analyze the computational complexity of the proposed method. For the fairness and comparability of the evaluation, commonly-used dataset\emph{ Princeton Shape Retrieval and Analysis Group}\footnote{\href{https://shape.cs.princeton.edu/benchmark/index.cgi}{https://shape.cs.princeton.edu/benchmark/index.cgi}}\cite{shilane2004princeton}  and 3D mesh objects: \emph{Beetle}, \emph{Mushroom}, \emph{Elephant}, \emph{Mannequin} shown in Fig.~\ref{fig_5} are adopted to test the validity of our method.
\subsection{Analysis of embedding rate }  
\begin{figure}[h]
  \centering
  \includegraphics[width=8.8cm,height=7cm]{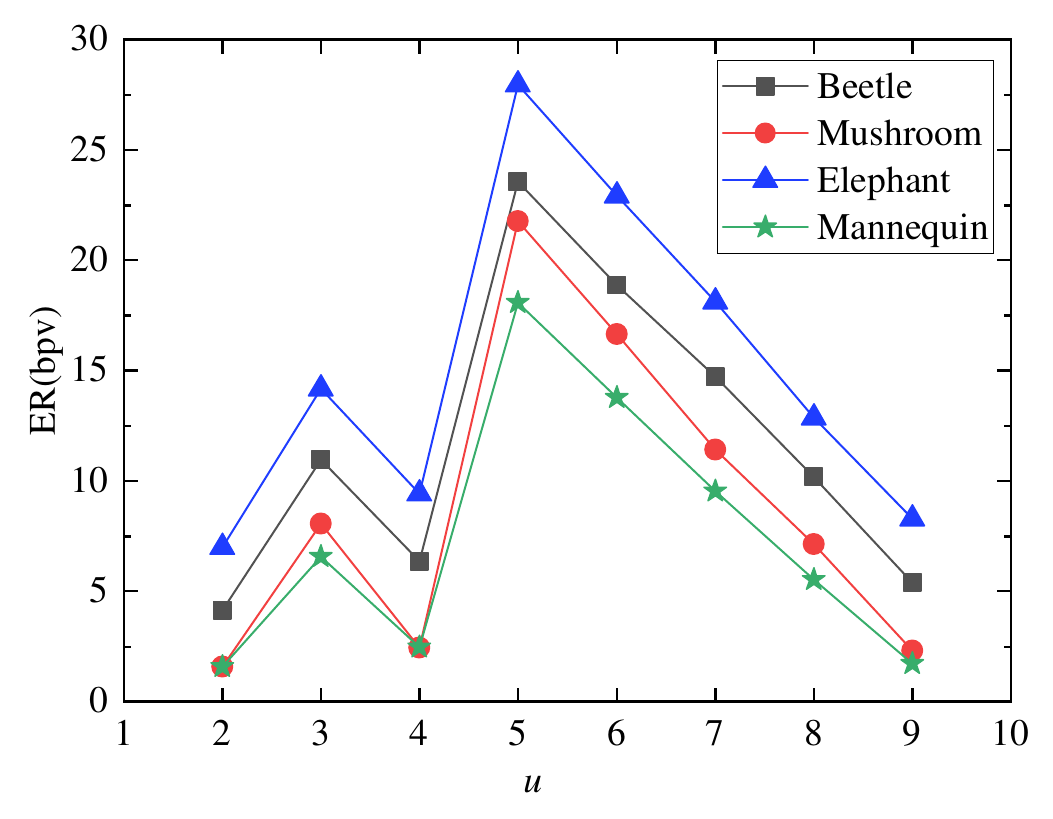}
\begin{center}
  \caption{ER (bpv) under different truncation coefficients $u$.}
  \label{fig_8}
\end{center}
\end{figure}
\begin{table*}
\setlength{\belowcaptionskip}{0.2cm}
\begin{center}
\renewcommand{\arraystretch}{1}
\caption{\centering Auxiliary information length and the ER (bpv) of different test 3D mesh models}
\label{tab_2}
\begin{tabular}{p{3cm}<{\centering}ccccp{1.8cm}<{\centering}}
\toprule[1pt]
Test models &Numbers of face &Numbers of vertices &Total EC (bit) & Label list length (bit) & ER (bpv) \\ 
\midrule
Beetle      &1763  & 988 & 25734          & 2466               & 23.55           \\ 
Mushroom    &448 & 226 & 5553          & 624                & 21.76           \\
Elephant    & 49918&24955&749634         & 51827               &27.96           \\
Mannequin   &839&428&8922         &1176              & 18.09         \\
\bottomrule[1pt]
\end{tabular}
\end{center}
\end{table*}
\par Bit per vertex (bpv) is generally used to describe the embedding rate of RDH in encrypted 3D mesh models. To illustrate embedding capacity of our method more clearly, The results of total embedding capacity (EC), the room occupied by the label list and maximum embedding rate in different test 3D mesh models are presented Table ~\ref{tab_2}. In our method, the embedding rate can be 
influenced by different truncation coefficient $u$. By selecting different $u$, we can obtain the optimal embedding rate. The correlation between the truncation coefficient $u$ and embedding rate is explored, as shown in Fig.~\ref{fig_8}.
In general, when $u$ is 1, the decimal reservation precision is too low to apply to most application scenarios and has no practical significance. Therefore, the proposed method takes the common precision of $u$ for study. Clearly, with the increase of $u$, the embedding rate shows an upward trend. The value of $u$ determines the bit length of coordinates. As the value of $u$ increases, so does the bit length of the coordinate, so more redundant room can be obtained. However, $u$ = 4, the embedding rate fluctuates greatly. According to Eq.(\ref{test_1}), when $u$ = 3 or 4, the vertex coordinates are represented by 16-bit bitstream. When $u$ = 4, compared with $u$ = 3, LSBs with lower correlation are introduced, embeddable room decreases and the embedding rate decreases significantly. When $u$ = 5, the vertex coordinates are converted to 32-bit bitstream and embeddable bits increase accordingly, so the embedding rate is significantly improved. When $u$ is greater than 5, more LSBs with low correlation are introduced, leading to the multi-MSB correlation of vertices decreases gradually, as does the embedding rate. 
\par Tagging the embeddable length of each vertex in our method makes full use of redundancy of vertex coordinates but generates auxiliary information. Arithmetic coding is used to compress label list with sparse features. This paper tests the difference of embedding rate before and after using arithmetic coding to compress label list as shown in Table ~\ref{tab_3}, which indicates that arithmetic coding can effectively reduce the room occupied by auxiliary information and reserve more room to embed additional data.
\begin{table*}[!t]
\setlength{\belowcaptionskip}{0.2cm}
\renewcommand{\arraystretch}{1.5}
\caption{\centering Comparison of ER (bpv) before and after arithmetic coding compression of auxiliary information}
\label{tab_3}
\centering
\begin{tabular}{ccm{0.9cm}<{\centering}m{0.9cm}<{\centering}m{0.9cm}<{\centering}m{0.9cm}<{\centering}m{0.9cm}<{\centering}m{0.9cm}<{\centering}m{0.9cm}<{\centering}m{0.9cm}<{\centering}}
\toprule[1pt]
\multirow{2}{*} & \multicolumn{9}{c}{$u$}   \\
\cmidrule{3-10}
\multicolumn{2}{c}{Test models}    & 2    & 3    & 4     & 5  & 6     & 7    & 8   & 9         \\ 
\cmidrule(r){1-10}  
                               & uncompressed & 3.75 & 10.20 & 5.61 & 23.08 & 18.49 & 14.02 & 9.37  & 4.66 \\  \cmidrule(r){2-2} \cmidrule(r){3-10} 
\multirow{-2}{*}{Beetle}   & \textbf{compressed}   & \textbf{4.13} & \textbf{10.97} & \textbf{6.35} & \textbf{23.55} & \textbf{18.88} & \textbf{14.73} & \textbf{10.17} & \textbf{5.38} \\ \cmidrule(r){1-10} 
                               & uncompressed & 1.04 & 7.57 & 2.19 & 21.57 & 16.39 & 10.97 & 7.04  & 1.89 \\\cmidrule(r){2-2} \cmidrule(r){3-10} 
\multirow{-2}{*}{Mushroom}  & \textbf{compressed}   & \textbf{1.59} & \textbf{8.05} & \textbf{2.45} & \textbf{21.76} &\textbf{16.65} & \textbf{11.42} & \textbf{7.13}  & \textbf{2.30} \\ \cmidrule(r){1-10}
                               &uncompressed & 6.39 & 13.82 & 8.95 & 27.03 & 22.16 & 17.31 & 12.45 & 7.52 \\\cmidrule(r){2-2} \cmidrule(r){3-10} 
\multirow{-2}{*}{Elephant}      & \textbf{compressed}   & \textbf{7.00} & \textbf{14.18} & \textbf{9.43} & \textbf{27.96} & \textbf{22.93} & \textbf{18.12} & \textbf{12.86} & \textbf{8.29} \\ \cmidrule(r){1-10}
                               & uncompressed &1.36 & 6.58  & 2.09 & 17.84 &13.64 & 9.19 & 5.17  & 0.96 \\ \cmidrule(r){2-2} \cmidrule(r){3-10}
\multirow{-2}{*}{Mannequin} &\textbf{compressed}   & \textbf{1.59} &\textbf{6.68}  & \textbf{2.48} & \textbf{18.05} & \textbf{13.78} & \textbf{9.55} & \textbf{5.53}  &\textbf{1.73} \\ 
\bottomrule[1pt]
\end{tabular}
\end{table*}

\begin{figure*}[!t]
\centering
\subfigure[]{
\centering
\begin{minipage}[b]{0.2\textwidth}
\includegraphics[width=1\textwidth]{figure/beetle.pdf} \\
\includegraphics[width=1\textwidth]{figure/mushroom.pdf}\\
\includegraphics[width=1\textwidth]{figure/elephant.pdf}\\
\includegraphics[width=1\textwidth]{figure/mannequin.pdf}
\end{minipage}
}
\subfigure[]{
\centering
\begin{minipage}[b]{0.22\textwidth}
\includegraphics[width=1\textwidth]{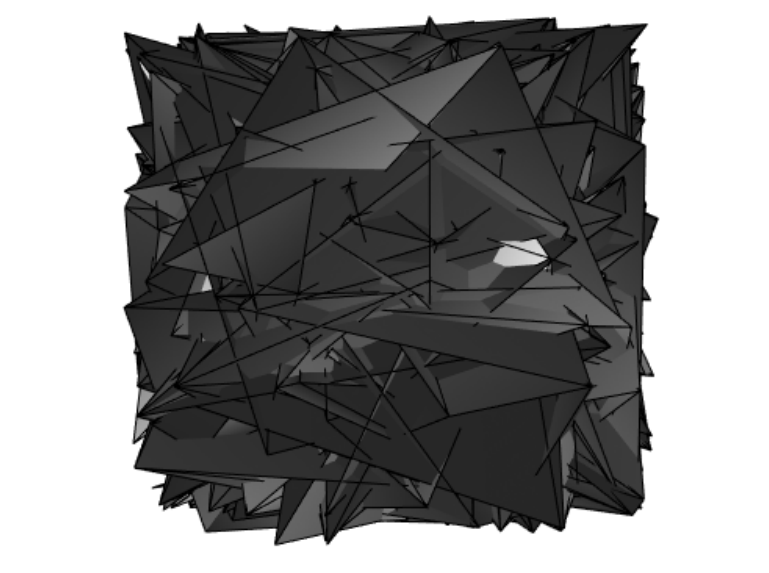} \\
\includegraphics[width=1\textwidth]{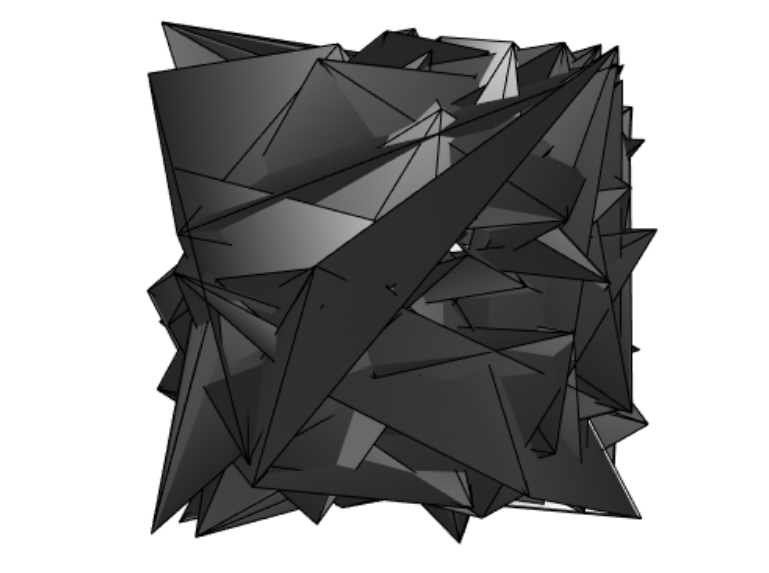}\\
\includegraphics[width=1\textwidth]{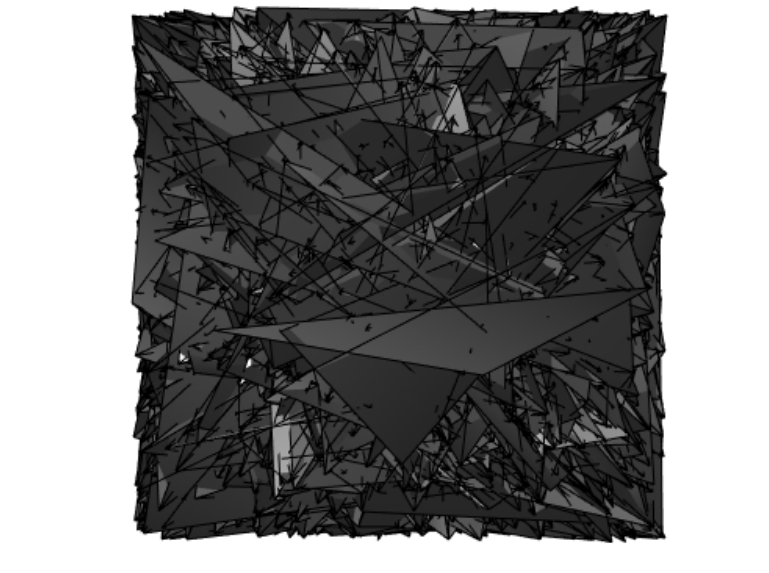}\\
\includegraphics[width=1\textwidth]{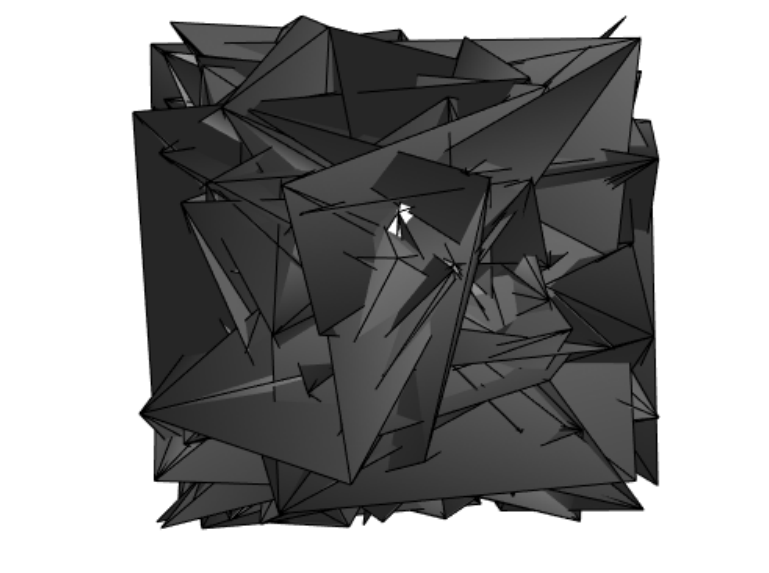}
\end{minipage}
}
\subfigure[]{
\begin{minipage}[b]{0.22\textwidth}
\includegraphics[width=1\textwidth]{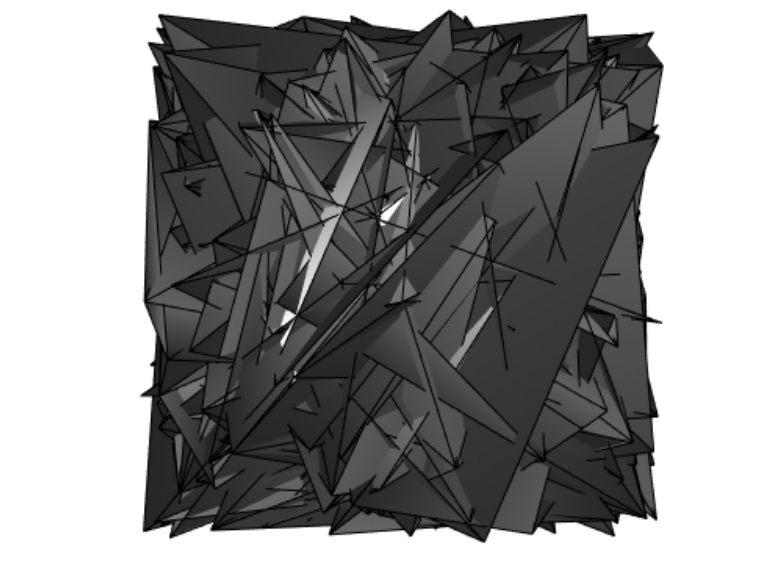} \\
\includegraphics[width=1\textwidth]{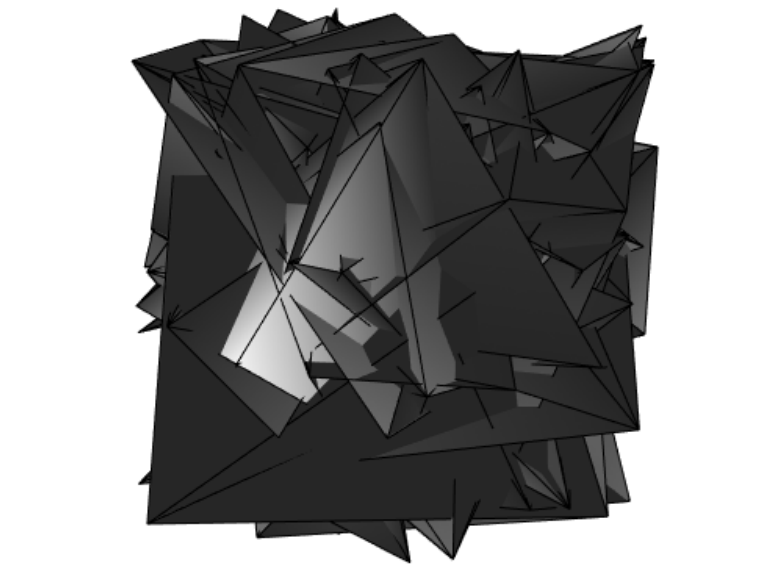}\\
\includegraphics[width=1\textwidth]{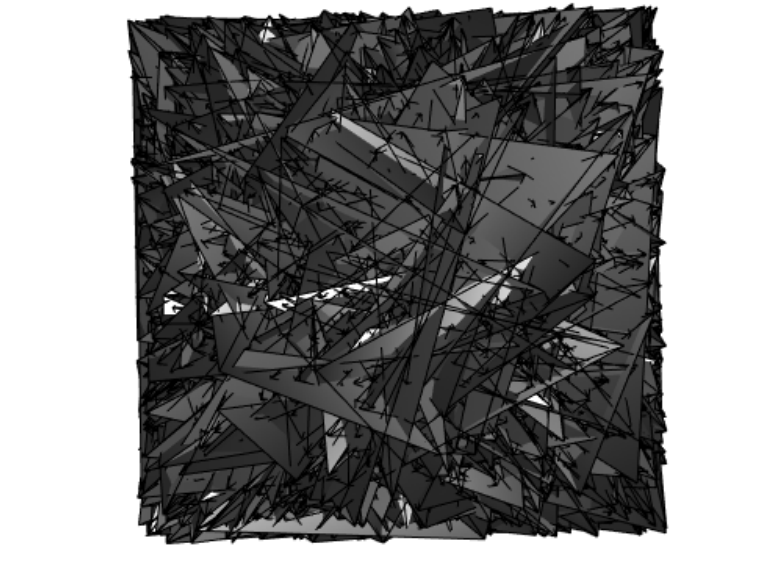}\\
\includegraphics[width=1\textwidth]{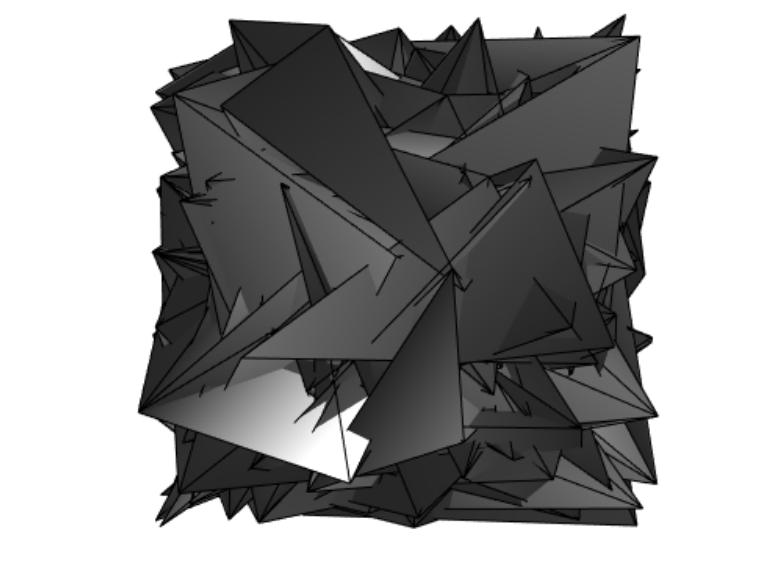}
\end{minipage}
}
\subfigure[]{
\begin{minipage}[b]{0.205\textwidth}
\includegraphics[width=1\textwidth]{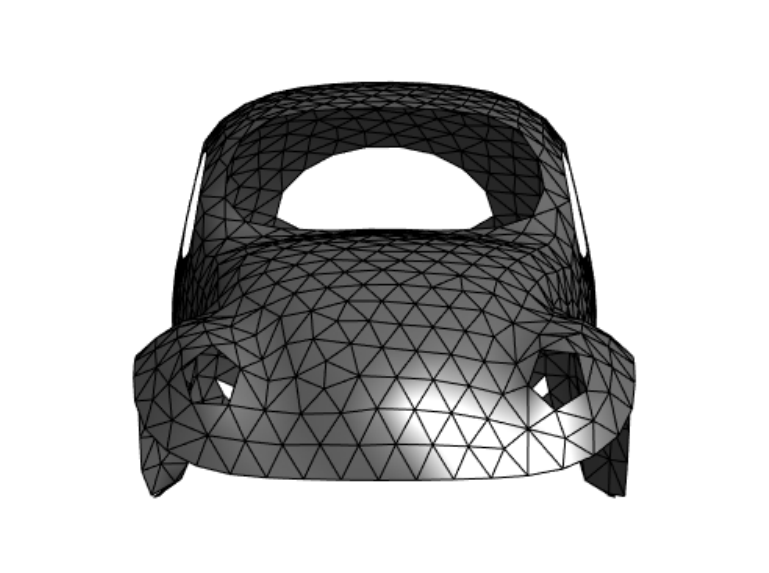} \\
\includegraphics[width=1\textwidth]{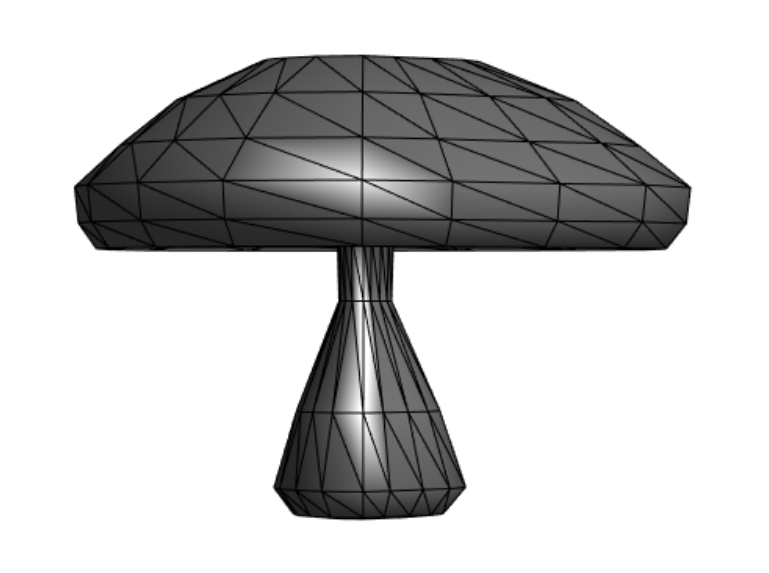}\\
\includegraphics[width=1\textwidth]{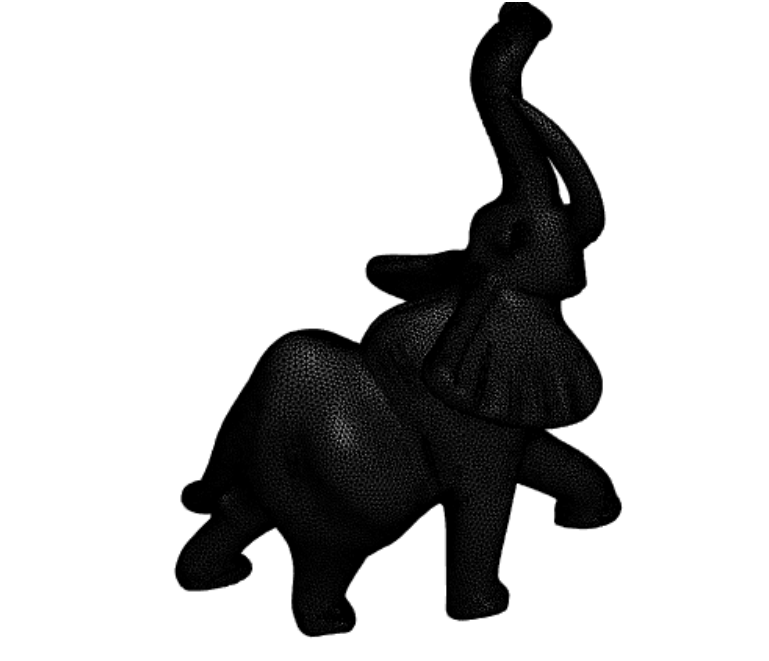}\\
\includegraphics[width=1\textwidth]{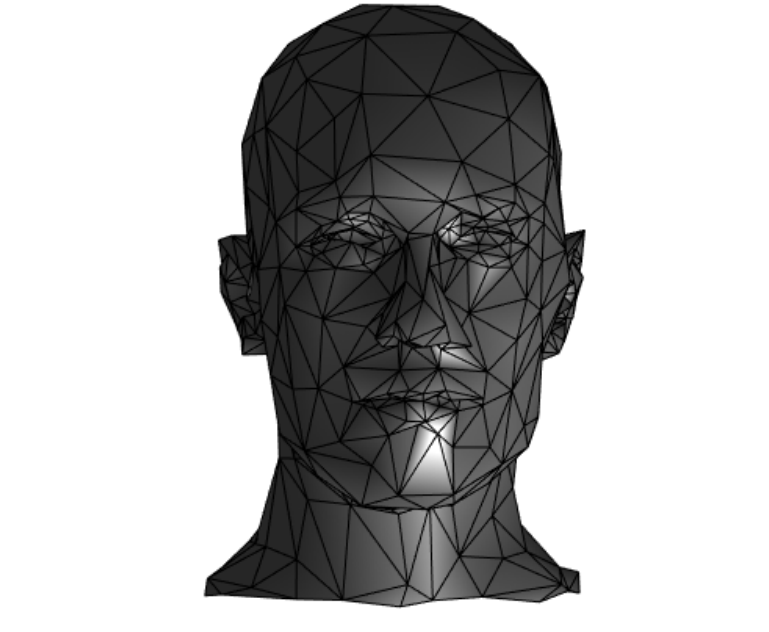}
\end{minipage}
}

\quad
\caption{\centering Visualization of four models change process: (a) original models; (b) encrypted models; (c) data-embedded models; (d) recovered models.}
\label{fig_9}
\end{figure*}

\begin{figure}[h]
\begin{center}
  \includegraphics[width=8.8cm,height=7cm]{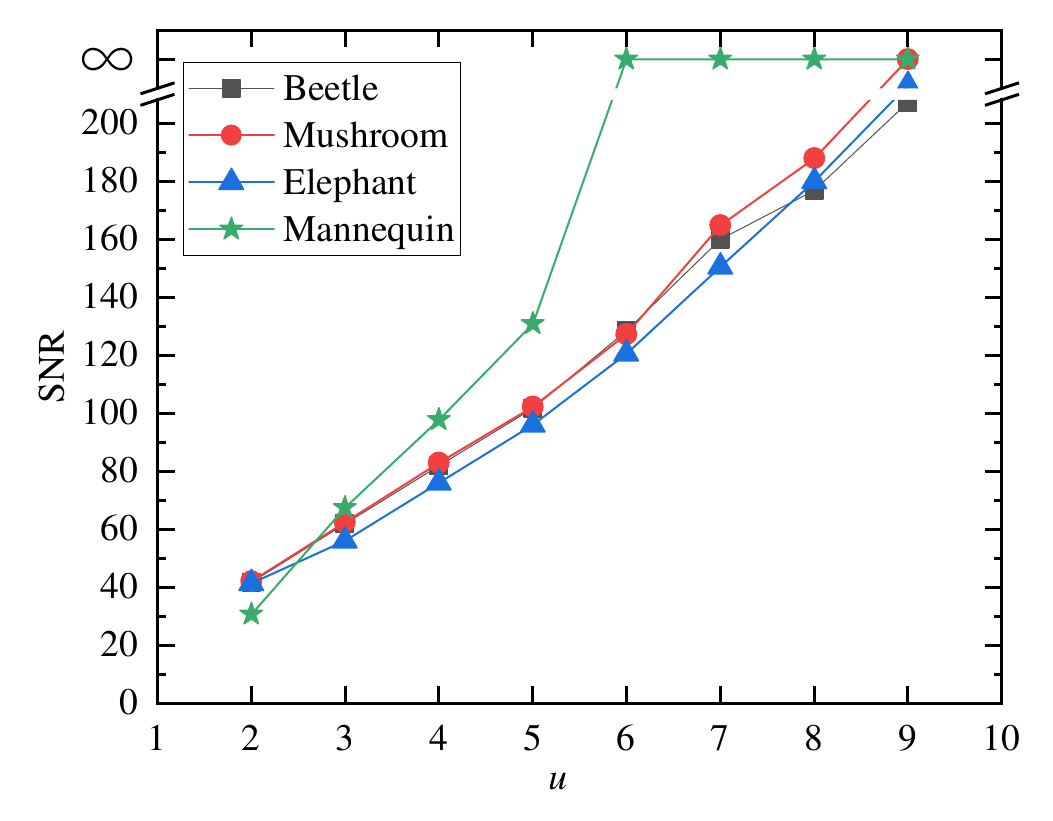}
  \caption{ SNR of four 3D mesh models with different truncation coefficient $u$ . }
  \label{fig_10}
\end{center}
\end{figure}

\begin{figure}[h]
\begin{center}
  \includegraphics[width=8.8cm,height=7cm]{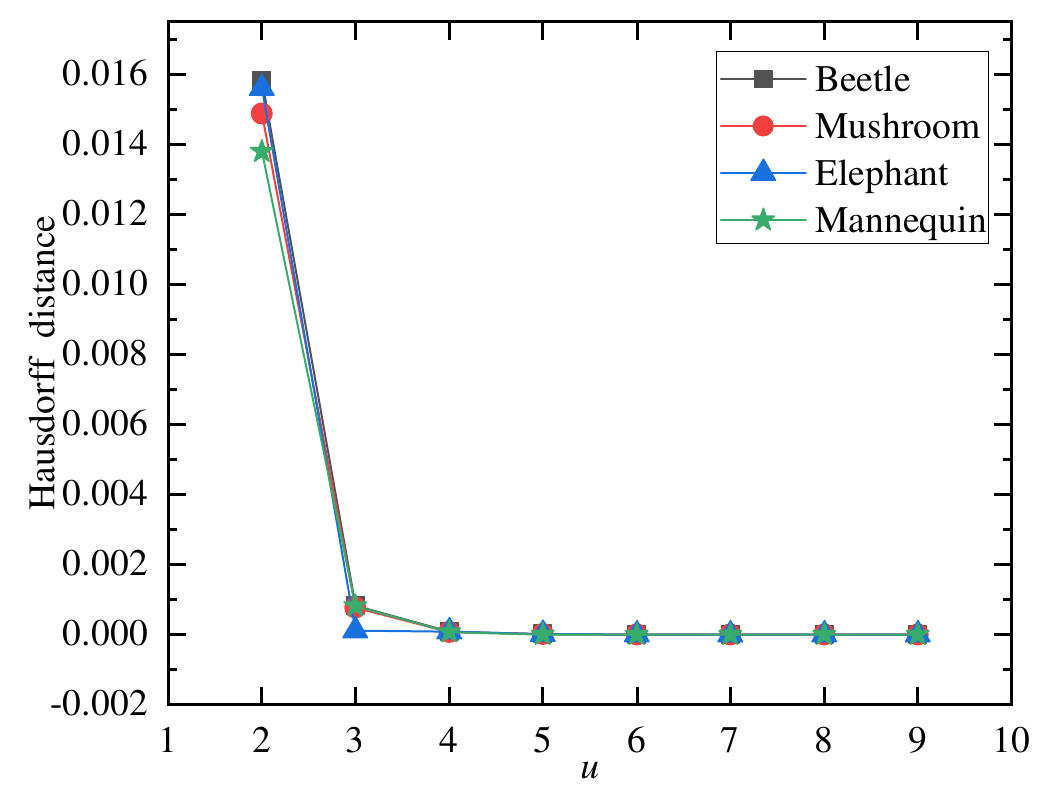}
  \caption{ Hausdorff distances between the recovered and the original 3D mesh models with different truncation coefficient $u$. }
  \label{fig_11}
\end{center}
\end{figure}
\subsection{Visual quality of restored model}
The appearance of four models at different phase, from left to right, is shown in Fig.~\ref{fig_9}. The differences between the reconstructed models and the original models cannot be found through the subjective perception of human eyes. To analyze the quality of model restoration more objectively, signal-to-noise ratio (SNR) and Hausdorff distance were introduced to evaluate the difference between the original model and the restored model~\cite{deering1995geometry}.
SNR can measure geometric distortion due to added noise, it can estimate the degree of vertex deviation of 3D model. The closer the SNR value is to $\infty$, the more similar the structure of the two models. SNR can be calculated by
\begin{eqnarray}
\begin{aligned}
&\text { SNR }= \\
& 10 \lg \frac{\sum_{i=1}^{N}\left[\left(v_{i, x}-\bar{v}_{x}\right)^{2}+\left(v_{i, y}-\bar{v}_{y}\right)^{2}+\left(v_{i, z}-\bar{v}_{z}\right)^{2}\right]}{\sum_{i=1}^{N}\left[\left(\hat{v}_{i, x}-v_{i, x}\right)^{2}+\left(\hat{v}_{i, y}-v_{i, y}\right)^{2}+\left(\hat{v}_{i, z}-v_{i, z}\right)^{2}\right]},
\end{aligned}
\end{eqnarray}
where $\bar{v}_{x}$, $\bar{v}_{y}$, $\bar{v}_{z}$ are the mean coordinates of the models; $v_{i, x}$, $ v_{i, y}$, $v_{i, z}$ are the original coordinates of $v_{i}$; and $\hat{v}_{i, x}$, $\hat{v}_{i, y}$, $\hat{v}_{i, z}$ are the modified coordinates of $v_{i}$. Hausdorff distance is the distance between two subsets of a metric space. The closer the Hausdorff distance is to 0, the smaller the difference between the two models. It can be represented by
\begin{eqnarray}
D(W,V)=\max _{w \in W}\left\{\min _{v \in V}\{d(w, v)\}\right\},
\end{eqnarray}
$W$ and $V$ are any two sets in Euclidean space, and $w$ and $v$ are two arbitrary points in these two sets respectively ~\cite{shah2018homomorphic}. Fig.~\ref{fig_10} shows the changing trend of SNR with different $u$. As $u$ increases, the SNR tends to $\infty$. Fig.\ref{fig_11} shows the Hausdorff distance value that measures the difference between the test model and its restored model with different $u$. With the increase of $u$, Hausdorff distance showed a decreasing trend. In other words, as the value of $u$ increases, the quality of model recovery is higher. This shows that $u$ have a great impact on the quality of model restoration. The reason for this is that the larger $u$ is, the more precise the vertex coordinates are. Therefore, when an appropriate $u$ is selected during coordinate transformation, the model can be restored with losslessly. In this paper, the mesh model can be reconstructed well when $u$ = 5 is chosen and the highest embedding rate is obtained, weighing the constraint between embedding rate and reconstruction quality.
\begin{figure*}[!t]
\begin{center}
  \includegraphics[width=14.8cm,height=8.4cm]{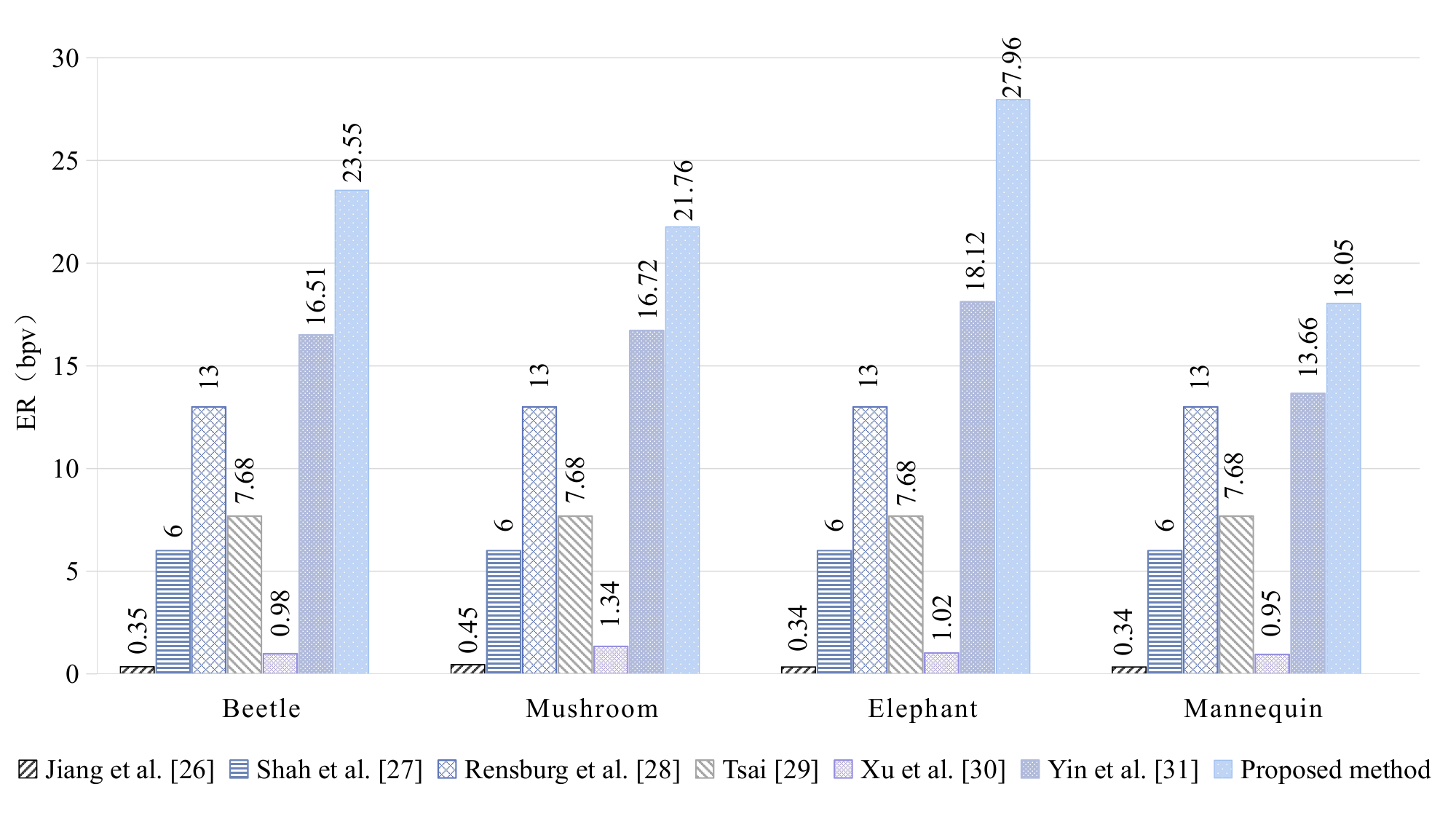}
  \caption{ Comparison of ER (bpv) of four test 3D mesh models. }
  \label{fig_6}
\end{center}
\end{figure*}
\begin{figure*}[!t]
\begin{center}
  \includegraphics[width=14.8cm,height=8.4cm]{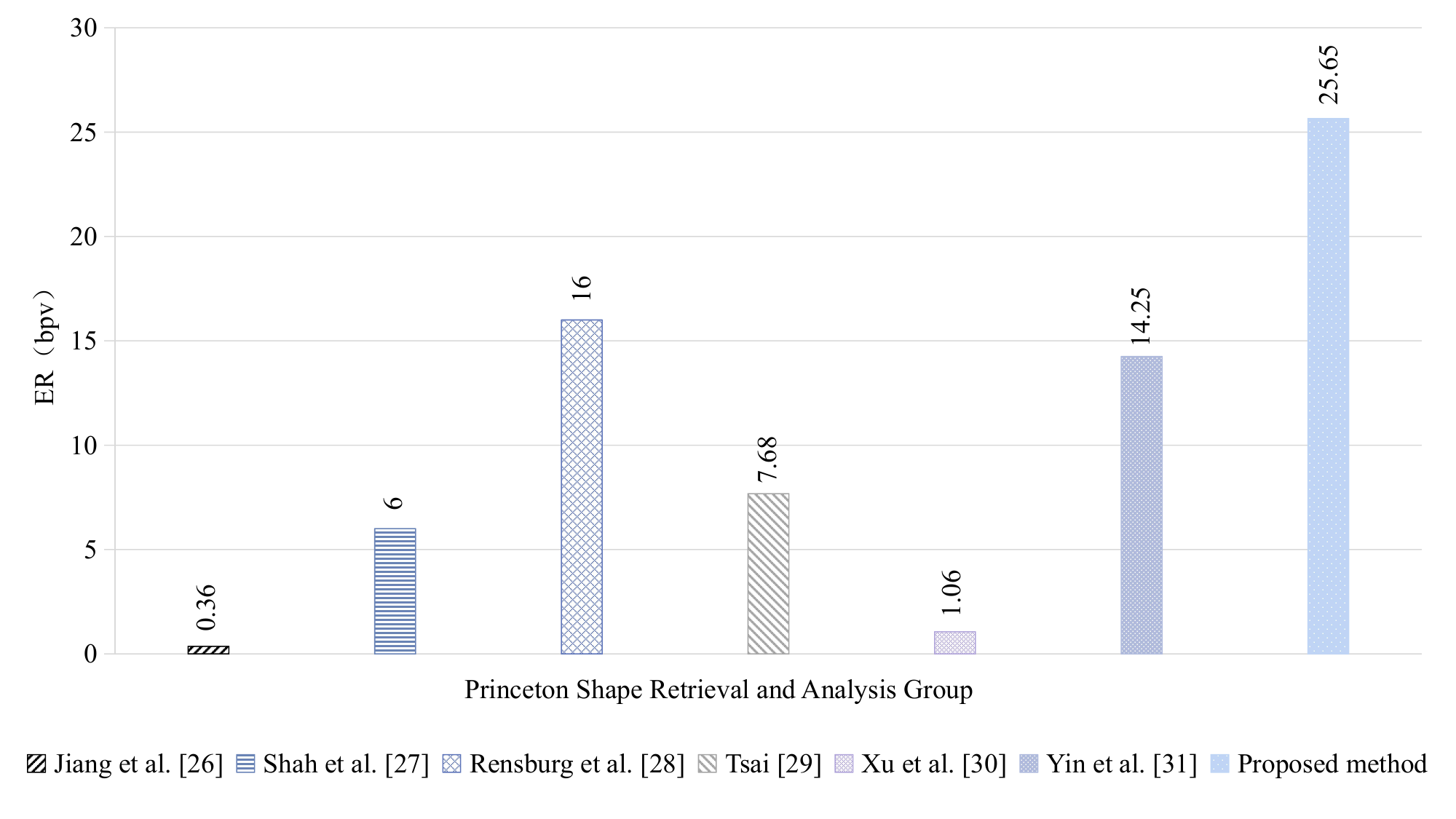}
  \caption{ Average ER (bpv) comparison in the Princeton Shape Retrieval and Analysis Group\cite{shilane2004princeton}. }
  \label{fig_7}
\end{center}
\end{figure*}
\subsection{Comparison with state-of-the-art methods}
\par We compare the proposed method performance with state-of-the-art works. The comparison results on the four test 3D objects of embedding rate are shown in Fig.~\ref{fig_6}. Furthermore, Fig.~\ref{fig_7} shows comparison of the maximum average embedding rate on dataset (\emph{Princeton Shape Retrieval and Analysis Group}\cite{shilane2004princeton}). Compared with ~\cite{yin2021separable}, the average embedding rate of the proposed method has improved by 11.40 bpv. In~\cite{jiang2017reversible} Jiang \emph{et al.} embed 1 bit data by flipping the LSBs of a portion of encrypted vertices and only part of vertices can embed data, therefore, the utilization of vertices is low, and the maximum embedding rate of this method is less than 1 bpv. Later, a two-layer embedding strategy was proposed in~\cite{shah2018homomorphic}. A two-layer embedding scheme is designed by using the histogram extension and the self-blind property of paillier cryptosystem, and the embedding rate reaches 6 bpv. Rensburg \emph{et al.}~\cite{van2021homomorphic} proposed a high capacity RDH-ED method based on homomorphic encryption. The method allows multiple bits data to be embedded in each vertex by formatting the model. In~\cite{tsai2020separable}, Tsai proposed an effective RDH method using spatial subdivision and space coding techniques. The spatial coding algorithm achieves high embedding rate close to the optimal algorithm. However, the embedding rate depends on the choice of embedding threshold. Choosing an inappropriate threshold may cause errors in data extraction. Unlike ~\cite{jiang2017reversible}, Xu \emph{et al.}~\cite{xu2021separable} exploit the strong correlation of the most significant bit to reserve space to improve the embedding rate to a certain extent. However, their method only used MSB to embed additional data, and not all vertex can be used for data embedding. Each vertex can embed no more than 3 bits of data. Therefore, there is room for improvement in embedding rate. Yin \emph{et al.}~\cite{yin2021separable} designed the method based on multi-MSB correlation. By checking the consistency between embeddable vertices and adjacent vertices from MSB to LSB bit values to obtain the maximum embeddable length. Their method obtained embedding capacity higher than previous work. The optimal embedding rate is explored by setting different embedding lengths. However, many vertices with embeddable lengths that do not reach the set length value are judged as vertices with wrong prediction and cannot be used for embedding data. In our method, half of the vertex coordinates can be embedded into the data and the embedding length of the vertex of ``embedded set'' is recorded by label list, which fully explores the embeddable room of vertices. Considering that the label list will occupy room, we use arithmetic coding to compress the label list. Experimental results prove that the proposed method can effectively improve the embedding capacity of the model.
\begin{table*}[!t]
\setlength{\belowcaptionskip}{0.2cm}
\renewcommand{\arraystretch}{1.3}
\caption{\centering Features comparison with state-of-the-art methods}
\label{tab_4}
\centering
\begin{tabular}{m{2cm}<{\centering}m{1.6cm}<{\centering}m{2cm}<{\centering}m{1.6cm}<{\centering}m{1.6cm}<{\centering}m{1.6cm}<{\centering}m{1.6cm}<{\centering}m{1.8cm}<{\centering}}
\toprule[1pt]
Features&Jiang et al.~\cite{jiang2017reversible}&Shah et al.~\cite{shah2018homomorphic}& Rensburg et al.~\cite{van2021homomorphic}&Tsai~\cite{tsai2020separable}&Xu et al.~\cite{xu2021separable}&Yin et al.~\cite{yin2021separable}&\textbf{Proposed\ method}\\
\midrule
Encryption\ method&Stream cipher&Homomorphic encryption&Homomorphic encryption& Stream cipher&Stream cipher&Stream cipher&\textbf{Stream cipher}\\
Computational\ complexity&Low&High&High&Low&Low&Low&\textbf{Low}\\
Error-free\ in\ data\ extraction&No&Yes&Yes&No&Yes&Yes&\textbf{Yes}\\
Error-free\ in\ model\ recovery &No&Yes&No&Yes&Yes&Yes&\textbf{Yes}\\
\bottomrule[1pt]
\end{tabular}
\end{table*}
\begin{table*}[!t]
\setlength{\belowcaptionskip}{0.2cm}
\renewcommand{\arraystretch}{1.5}
\centering
\caption{Run-time of the proposed method in four test 3D mesh models}
\begin{tabular}{m{4cm}<{\centering}m{3.8cm}<{\centering}m{3.8cm}<{\centering}m{3.8cm}<{\centering}} 
\toprule[1pt]
Models  & Numbers of face & Numbers of vertices  & Run-time in seconds \\
        \midrule
Beetle & 1763 & 988 & 14.36 \\
Mushroom&448 &226 &3.61\\
Elephant&49918  & 24955 & 1461.42 \\
Mannequin &839 &428 &6.19\\
  \bottomrule[1pt]
  \end{tabular}
  \label{tab_5}
\end{table*}
\par Table ~\ref{tab_4} is the results of the proposed method compared with state-of-the-art methods in terms of features. In~\cite{jiang2017reversible}, vertex coordinates are converted into binary forms of corresponding lengths according to rules. By flipping the LSBs, the additional data 0 or 1 is embedded into the encrypted vertices. When extracting data, the smooth function measuring fluctuation is used to locate the change of the local area of the model to find the position of data embedding. However, there are errors in using smooth function to evaluate embedded data. In~\cite{shah2018homomorphic}, Shah \emph{et al.} proposed embedding additional data by self-blinding property of Paillier encryption. Based on this property, embedding data in the encrypted domain does not affect the plaintext value, thus extracting data accurately and restoring the original model without distortion. However, 3D mesh models encrypted with a Pailliar homomorphic cryptosystem will cause vertex coordinate value expansion and high computational complexity which limits its practical application. In~\cite{van2021homomorphic}, a two-layer RDH-ED method based on homomorphic encryption is proposed. To enhance security, vertices are first grouped into blocks. Vertex data expansion is addressed by setting the specified bit value in the block to 0. However, homomorphic encryption results in low computational efficiency and the method cannot recover the original model losslessly.
Tsai~\cite{tsai2020separable} designed a RDH method based on 3D models by using spatial subdivision and space coding techniques. First, one determines a ratio based on the distance between the vertex and the minimum boundary point and the side length of the bounding volume and encrypts the ratio already obtained. Using XOR operation to encrypt model improves the efficiency of encryption. However, the correctness of data extraction depends on the threshold value, if the threshold is not selected properly, some errors may occur when extracting data. In~\cite{xu2021separable} and~\cite{yin2021separable}, bit-substitution is used to embed additional data, which can extract data accurately. The model can be restored losslessly by choosing different truncation coefficients. In our method, stream cryptography is used to encrypt the 3D models, which improves the efficiency of the method. On the premise of ensuring the prediction accuracy, the vertex set was divided and the correlation of neighborhood vertices was used to restore the models. From the above experimental results, it can be seen that the proposed method can accurately extract additional data, and the original model can be restored losslessly by selecting appropriate truncation coefficient.
\subsection{Analysis of computational complexity}
The main phases included in the proposed method are the process of model encryption, data embedding, and model recovery. A 3D mesh model with $M$ triangular face and $N$ vertices, the operation of model encrypt phase is to traverse all vertices and convert them into binary form of corresponding length according to vertex coordinate transformation rule, the time consumption of this phase is $O(3N)$. In data embedding phase, data is embedded by bit substitution, the time cost in this phase is $O(M+3/2N)$. In the final phase, all vertices need to be classified as ``prediction set'' and ``embedded set'', and then data is extracted from the ``embedded set'' vertices. Finally, all vertices  are recovered by using the correlation of 1-ring neighborhood vertices, the time complexity is $O(M+3/2N +3MN)$. The run-time is tested on four models, and the results are shown in Table ~\ref{tab_5}. It can be seen that the run-time of the complex model is longer due to the large number of vertices traversed, while that of the simple model is shorter. 

\section{Conclusion}
\label{sec::Conclusion}
\par This paper proposes a high-capacity RDH method in encrypted 3D mesh models method based on multi-MSB prediction. To better balance vertices utilization and vertices correction accuracy, The method divides vertices into ``embedded set'' for embedding data and ``prediction set'' for recovering vertices based on the odd-even property of indices. Self-embedding data according to the label map, making full use of the redundant bits per vertex. Arithmetic coding compresses auxiliary information to further free up effective room. The model vertices can be recovered without loss by exploiting the correlation of 1-ring neighborhood vertices. Experimental results prove the effectiveness of the proposed method, and the embedding rate of the proposed method outperforms the state-of-the-art methods.
\par In follow-up research, we will focus on improving the robustness of encrypted model against external attacks. In addition, for complex models, we will further improve the algorithm to save the vertex traversal time.

\section*{Acknowledgment}
This research work is partly supported by National Natural Science Foundation of China (62172001,61872003).

\bibliography{mybibfile}

\begin{thebibliography}{10}
\expandafter\ifx\csname url\endcsname\relax
  \def\url#1{\texttt{#1}}\fi
\expandafter\ifx\csname urlprefix\endcsname\relax\def\urlprefix{URL }\fi
\expandafter\ifx\csname href\endcsname\relax
  \def\href#1#2{#2} \def\path#1{#1}\fi

\bibitem{barton1997method}
J.~M. Barton, Method and apparatus for embedding authentication information
  within digital data, United States Patent, 5 646 997 (1997).

\bibitem{puech2008reversible}
W.~Puech, M.~Chaumont, O.~Strauss, A reversible data hiding method for
  encrypted images, in: Security, forensics, steganography, and watermarking of
  multimedia contents X, Vol. 6819, SPIE, 2008, pp. 534--542.

\bibitem{zhang2011reversible}
X.~Zhang, Reversible data hiding in encrypted image, IEEE signal processing
  letters 18~(4) (2011) 255--258.

\bibitem{ni2006reversible}
Z.~Ni, Y.-Q. Shi, N.~Ansari, W.~Su, Reversible data hiding, IEEE Transactions
  on circuits and systems for video technology 16~(3) (2006) 354--362.

\bibitem{jia2019reversible}
Y.~Jia, Z.~Yin, X.~Zhang, Y.~Luo, Reversible data hiding based on reducing
  invalid shifting of pixels in histogram shifting, Signal Processing 163
  (2019) 238--246.

\bibitem{gao2020reversible}
X.~Gao, Z.~Pan, E.~Gao, G.~Fan, Reversible data hiding for high dynamic range
  images using two-dimensional prediction-error histogram of the second time
  prediction, Signal Processing 173 (2020) 107579.

\bibitem{tian2003reversible}
J.~Tian, Reversible data embedding using a difference expansion, IEEE
  transactions on circuits and systems for video technology 13~(8) (2003)
  890--896.

\bibitem{fan2021reversible}
G.~Fan, Z.~Pan, E.~Gao, X.~Gao, X.~Zhang, Reversible data hiding method based
  on combining ipvo with bias-added rhombus predictor by multi-predictor
  mechanism, Signal Processing 180 (2021) 107888.

\bibitem{celik2006lossless}
M.~U. Celik, G.~Sharma, A.~M. Tekalp, Lossless watermarking for image
  authentication: a new framework and an implementation, IEEE Transactions on
  Image Processing 15~(4) (2006) 1042--1049.

\bibitem{zhang2013recursive}
W.~Zhang, X.~Hu, X.~Li, N.~Yu, Recursive histogram modification: establishing
  equivalency between reversible data hiding and lossless data compression,
  IEEE transactions on image processing 22~(7) (2013) 2775--2785.

\bibitem{wu2005reversible}
H.-T. Wu, Y.-M. Cheung, A reversible data hiding approach to mesh
  authentication, in: The 2005 IEEE/WIC/ACM International Conference on Web
  Intelligence (WI'05), IEEE, 2005, pp. 774--777.

\bibitem{wu2008reversible}
H.-T. Wu, J.-L. Dugelay, Reversible watermarking of 3d mesh models by
  prediction-error expansion, in: 2008 IEEE 10th workshop on multimedia signal
  processing, IEEE, 2008, pp. 797--802.

\bibitem{zhang2019reversibility}
Q.~Zhang, X.~Song, T.~Wen, C.~Fu, Reversibility improved data hiding in 3d mesh
  models using prediction-error expansion and sorting, Measurement 135 (2019)
  738--746.

\bibitem{jiang2018reversible}
R.~Jiang, W.~Zhang, D.~Hou, H.~Wang, N.~Yu, Reversible data hiding for 3d mesh
  models with three-dimensional prediction-error histogram modification,
  Multimedia Tools and Applications 77~(5) (2018) 5263--5280.

\bibitem{luo2006reversible}
H.~Luo, Z.-M. Lu, J.-S. Pan, A reversible data hiding scheme for 3d point cloud
  model, in: 2006 IEEE International Symposium on Signal Processing and
  Information Technology, IEEE, 2006, pp. 863--867.

\bibitem{sun2006reversible}
Z.~Sun, Z.-M. Lu, Z.~Li, Reversible data hiding for 3d meshes in the
  pvq-compressed domain, in: 2006 international conference on intelligent
  information hiding and multimedia, IEEE, 2006, pp. 593--596.

\bibitem{lu2007high}
Z.-M. Lu, Z.~Li, High capacity reversible data hiding for 3d meshes in the pvq
  domain, in: International Workshop on Digital Watermarking, Springer, 2007,
  pp. 233--243.

\bibitem{bhardwaj2022efficient}
R.~Bhardwaj, Efficient separable reversible data hiding algorithm for
  compressed 3d mesh models, Biomedical Signal Processing and Control 73 (2022)
  103265.

\bibitem{zhang2011separable}
X.~Zhang, Separable reversible data hiding in encrypted image, IEEE
  transactions on information forensics and security 7~(2) (2011) 826--832.

\bibitem{wu2014high}
X.~Wu, W.~Sun, High-capacity reversible data hiding in encrypted images by
  prediction error, Signal processing 104 (2014) 387--400.

\bibitem{hong2012improved}
W.~Hong, T.-S. Chen, H.-Y. Wu, An improved reversible data hiding in encrypted
  images using side match, IEEE signal processing letters 19~(4) (2012)
  199--202.

\bibitem{qin2018separable}
C.~Qin, W.~Zhang, F.~Cao, X.~Zhang, C.-C. Chang, Separable reversible data
  hiding in encrypted images via adaptive embedding strategy with block
  selection, Signal Processing 153 (2018) 109--122.

\bibitem{ma2013reversible}
K.~Ma, W.~Zhang, X.~Zhao, N.~Yu, F.~Li, Reversible data hiding in encrypted
  images by reserving room before encryption, IEEE Transactions on information
  forensics and security 8~(3) (2013) 553--562.

\bibitem{zhang2014reversibility}
W.~Zhang, K.~Ma, N.~Yu, Reversibility improved data hiding in encrypted images,
  Signal Processing 94 (2014) 118--127.

\bibitem{yin2021reversible}
Z.~Yin, X.~She, J.~Tang, B.~Luo, Reversible data hiding in encrypted images
  based on pixel prediction and multi-msb planes rearrangement, Signal
  Processing 187 (2021) 108146.

\bibitem{jiang2017reversible}
R.~Jiang, H.~Zhou, W.~Zhang, N.~Yu, Reversible data hiding in encrypted
  three-dimensional mesh models, IEEE Transactions on Multimedia 20~(1) (2017)
  55--67.

\bibitem{shah2018homomorphic}
M.~Shah, W.~Zhang, H.~Hu, H.~Zhou, T.~Mahmood, Homomorphic encryption-based
  reversible data hiding for 3d mesh models, Arabian Journal for Science and
  Engineering 43~(12) (2018) 8145--8157.

\bibitem{van2021homomorphic}
B.~J. van Rensburg, P.~Puteaux, W.~Puech, J.-P. Pedeboy, Homomorphic two tier
  reversible data hiding in encrypted 3d objects, in: 2021 IEEE International
  Conference on Image Processing (ICIP), IEEE, 2021, pp. 3068--3072.

\bibitem{tsai2020separable}
Y.-Y. Tsai, Separable reversible data hiding for encrypted three-dimensional
  models based on spatial subdivision and space encoding, IEEE Transactions on
  Multimedia (2020).

\bibitem{xu2021separable}
N.~Xu, J.~Tang, B.~Luo, Z.~Yin, Separable reversible data hiding based on
  integer mapping and msb prediction for encrypted 3d mesh models, Cognitive
  Computation (2021) 1--10.

\bibitem{yin2021separable}
Z.~Yin, N.~Xu, F.~Wang, L.~Cheng, B.~Luo, Separable reversible data hiding
  based on integer mapping and multi-msb prediction for encrypted 3d mesh
  models, in: Chinese Conference on Pattern Recognition and Computer Vision
  (PRCV), Springer, 2021, pp. 336--348.

\bibitem{deering1995geometry}
M.~Deering, Geometry compression, in: Proceedings of the 22nd annual conference
  on Computer graphics and interactive techniques, 1995, pp. 13--20.

\bibitem{shilane2004princeton}
P.~Shilane, P.~Min, M.~Kazhdan, T.~Funkhouser, The princeton shape benchmark,
  in: Proceedings Shape Modeling Applications, 2004., IEEE, 2004, pp. 167--178.

\end{thebibliography}

\vspace{2em}
\par{
  \begin{wrapfigure}{l}{20mm}
    \includegraphics[width=1in,height=1in,clip,keepaspectratio]{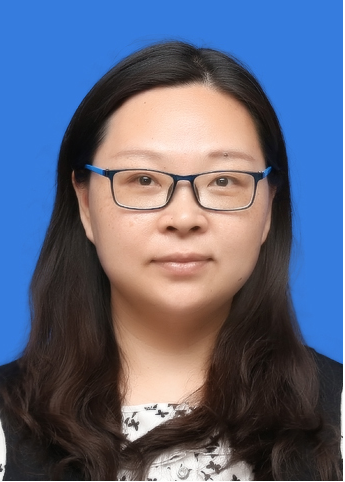}
  \end{wrapfigure}\par
  \textbf{Wan-Li Lyu} received the M.S. degree in computer science and technology with Guangxi University and the Ph.D. degree in computer science and technology with Anhui University. She was a postdoctoral research fellow in Department of Information Engineering and Computer Science at Feng Chia University from August 2013 to July 2014. Since July 2004, she is a Lecturer in School of Computer Science and Technology, Anhui University. Her current research interests include image processing, data hiding and information security.\par}

\vspace{2em}
\par{
  \begin{wrapfigure}{l}{20mm}
    \includegraphics[width=1in,height=1in,clip,keepaspectratio]{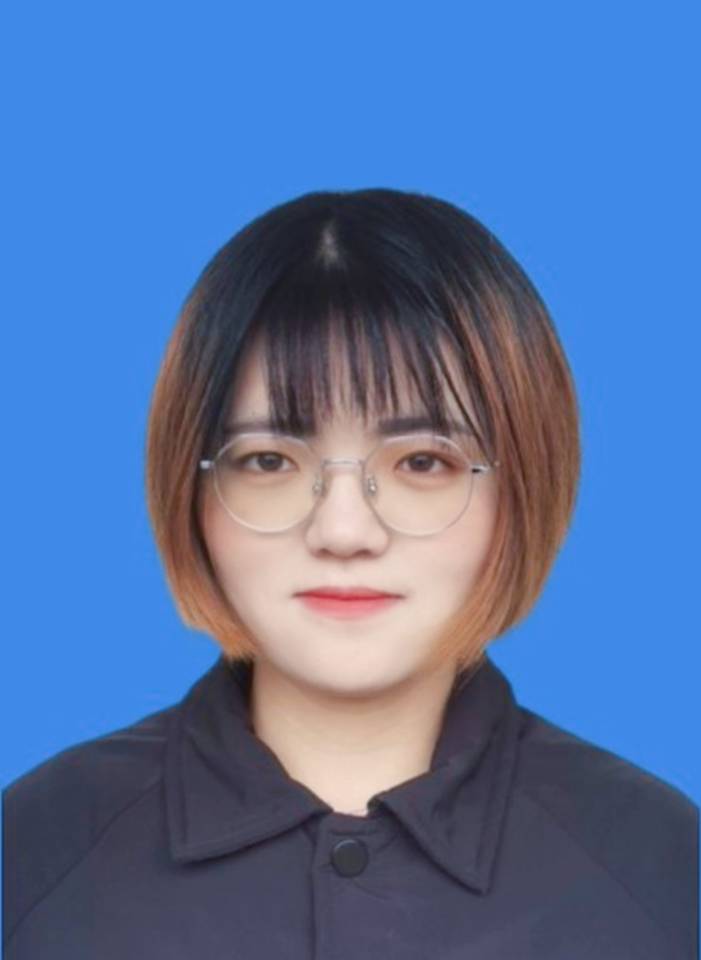}
  \end{wrapfigure}\par
  \textbf{Lulu Cheng} received her bachelor degree in Computer Science and Technology in 2019 and now is a master student in the School of Computer Science and Technology, Anhui University. Her current research interests include reversible data hiding in encrypted domain.\par}

\vspace{2em}
\par{
  \begin{wrapfigure}{l}{20mm}
    \includegraphics[width=1in,height=1in,clip,keepaspectratio]{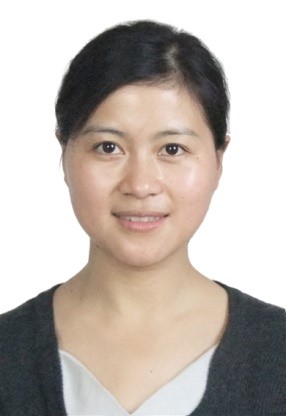}
  \end{wrapfigure}\par
  \textbf{Zhaoxia Yin} received her B.Sc., M.E. \& Ph.D. from Anhui University in 2005, 2010, and 2014 respectively. She is a senior member of CSIG and Digital Media Forensics and Security Professional Committee. She is also a IEEE/ACM/CCF member and served CCF YOCSEF Hefei as an Associate Chair of the academic committee from 2016-2017. She was an Associate Professor and Doctoral Tutor in School of Computer Science and Technology at Anhui University. Currently she works as a full propfesser in School of Communication \& Electronic Engineering at East China Normal University. Her primary research focus including Data Hiding, Privacy \& Security of Multimedia \& Machine Learning and she is the Principal Investigator of three NSFC Projects.\par}
\end{document}